\documentclass[article]{aa}
\usepackage[utf8]{inputenc}
\usepackage{txfonts}
\usepackage{xcolor} 
\usepackage{ulem} 

\def\vec#1{\mathbf{#1}}

\title{Planetesimal and planet formation in transient dust traps}
\author{ Zs. S\'andor
          \inst{1,2,3}
          \and
	   O. M. Guilera
          \inst{4,5}
          \and 
           Zs. Reg\'aly 
          \inst{2,3}
	  \and
	   W. Lyra
	  \inst{6}}
	  
	  \institute{
             ELTE E\"otv\"os Lor\'and University, Institute of Physics and Astronomy, Department of Astronomy,
             H-1117 Budapest, P\'azm\'any P\'eter s\'et\'any 1/A, Hungary. \email{Zs.Sandor@astro.elte.hu}
         \and
             Konkoly Observatory, Research Centre for Astronomy and Earth Sciences, H-1121 Budapest, Konkoly Thege Miklós út 15-17, Hungary. \email{regaly.zsolt@csfk.org}
        \and
             CSFK, MTA Centre of Excellence, Budapest, Konkoly Thege Miklós út 15-17., H-1121, Budapest, Hungary
         \and
             Astrophysical Institute of La Plata, National Scientific and Technical Research Council and National University of La Plata, Paseo del Bosque s/n, 1900 La Plata, Argentina. \email{oguilera@fcaglp.unlp.edu.ar}
         \and
             Millennium Nucleus of Planetary Formation (NPF), Chile. 
         \and 
             Department of Astronomy, New Mexico State University, Las Cruces, New Mexico, USA. \email{wlyra@nmsu.edu}
         }
\date{December, 2023}

\abstract{The ring-like structures in protoplanetary discs that are observed in the cold dust emission by ALMA, might be explained by dust aggregates trapped aerodynamically in pressure maxima.
}{The effect of a transient pressure maximum is investigated that develops between two regimes with different turbulent levels. We study how such a pressure maximum collects dust aggregates and transforms them into large planetesimals and Moon-mass cores that can further grow to a few Earth-mass planets by pebble accretion, and eventually to giant planets, by considering the accretion of a gaseous envelope.  
}{A numerical model is developed, incorporating the evolution of gaseous disc, growth and transport of pebbles, N-body interactions of growing planetary cores and their backreaction to gas disc by opening a partial gap. Planetesimal formation by streaming instability is parametrized in our model.
}{A transient pressure maximum efficiently accumulates dust particles that can grow larger than mm-size. If this happens, dust aggregates can be transformed by the streaming instability process into such large planetesimals, which can grow further by pebble accretion, according to our assumptions. As the gas evolves to its steady state, the pressure maximum vanishes, and the concentrated pebbles that are not transformed to planetesimals and accreted by the growing planet, drift inward. During this inward drift, if the conditions of the streaming instability are met, planetesimals are formed in a wide radial range of the disc.
}{A transient pressure maximum is a favourable place for planetesimal and planet formation during its lifetime and the concentration of pebbles induces continuous formation of planetesimals even after its disappearance. Besides, the formation of a planet can trigger the formation of planetesimals over a wide area of the protoplanetary disc.} 

    
\begin{document}

\maketitle

\section{Introduction}
The concept of planet formation in pressure maxima of protoplanetary discs has been proposed to avoid the quick loss and destruction of solid material by rapid inward drift and high-velocity collisions, a problem commonly known as the ``meter-size barrier'' \citep{BlumWurm2008}.
As a possible solution, it has been assumed that there are preferential places for dust and planet growth where the aerodynamical drag felt by dust particles and torques that lead to planet migration vanish or significantly decrease. These places are named dust and planet traps, respectively. Particularly interesting is when a density maximum generates a planet trap, since a pressure maximum may also develop in connection. Pressure maxima act as dust traps, collecting grains, and helping their further growth to planetesimal, or even to embryo sizes via either coagulation \citep{Brauer+2008}, or drag-induced instabilities \citep{YoudinShu2002ApJ, Johansen+2009}. If due to the subsequent accretion processes massive embryos form in a pressure maximum, they might be locked in the nearby developed planet trap. The combined effect of a planet trap and pressure maximum in the formation of larger bodies, with sizes ranging between Mars and Jupiter, have been investigated by \citet{Lyra+2008, Lyra+2009, Sandor+2011ApJ, Regaly+2013MNRAS, GuileraSandor2017A&A, Guilera+2020, Morbidelli2020A&A, Chambers2021ApJ}, and \citet{SandorRegaly2021MNRAS} among many others.

Pressure traps are naturally expected to develop in disc locations where accretion flow transitions occur. An ideal candidate for such a transition could be the boundary between an accretionally active and inactive region, in which the magnetorotational instability (MRI) plays the central role \citep{BalbusHawley1991ApJ}, and of a region where the low ionization fraction of disc renders the gas accreting in a residual level. On the other hand, the role of the MRI in protoplanetary discs has been questioned in the past decade. As recent results indicate, the structure of the dead zone is quite complex: the Ohmic dead zone, whose existence was suggested by \cite{Gammie1996ApJ}, is followed by a region in which the Hall effect is dominant being prone to Hall shear instability, and finally, a region where ambipolar diffusion is the most important effect that by suppressing the MRI, turns easier the magnetic breaking by disc wind \citep{Lesur+2014A&A}, see a recent review of \citet{LyraUmurhan2019PASP}. Despite these new results in the understanding of the outer disc's ionization structure, the existence of a transition between regions with different accretion strengths is still generally assumed \citep{Delage+2022A&A, Lesur+2022}. It is important to note that if the accretionally weaker disc region is followed by an accretionally stronger one, both the gas surface density maximum and the pressure maximum developed at the boundary of these regions vanish as the disc adapts to its steady state \citep{GuileraSandor2017A&A}. For this reason, such maxima have limited lifetimes and can be considered transient ones. 

A key element of our proposed scenario is the development of such transient surface density/pressure maxima of the disc's gas since the high concentration of dust grains in pressure maxima may lead to planetesimal formation via the streaming instability (SI), being a mechanism first proposed by \citet{YoudinGoodman2005}. Investigations of \cite{Carrera2021} show that SI is triggered for dust aggregates with cm-size in pressure bumps. Later, \citet{CarreraSimon2022} found that triggering the SI in pressure bumps may be problematic for dust particles in the mm-size regime being most accessible to (sub) millimetre observation with ALMA. These observations revealed ring structures of dust in protoplanetary discs that are associated with pressure maxima \citep[e.g][]{Dullemond2018} being considered as preferential places of planet formation. Nevertheless, the formation of planetesimals by SI or by direct gravitational collapse in such pressure maxima is still an open problem being the subject of intensive research \citep[e.g][]{Carrera2021, CarreraSimon2022, XuBai2022a, XuBai2022b}.

In the case of a static pressure maximum, if the SI is triggered, planetesimal formation is localized only to the vicinity of the pressure maximum. This is contradicted by the fact that planetesimals are distributed throughout the disc, as Solar System examples (main belt of asteroids, Kuiper belt objects, etc) suggest. 

There are other places, where planetesimal formation can be triggered. For instance, at the water snowline the disc's solid material can be piled up: inside the snowline, water ice evaporates, however by outer diffusion water vapour can penetrate to the low-temperature region (beyond the snowline), where by freezing out to the solid dust aggregates increases the surface density of the solid material \citep[see e.g.][]{StevensonLunine1988Icar}. This enhancement of the solid particles has been investigated in the context of planetesimal formation more recently by \citet{DrazkowskaAlibert2017A&A}. In one of their initially high metallicity disc models ($Z=0.05$), the outer boundary of planetesimals can be extended up to 16 AU, however, in the lower metallicity (and more realistic) cases the outer boundary extends up to 7 AU. In addition, \citet{DrazkowskaDullemond2018A&A, DrazkowskaDullemond_corr2023A&A} showed that the migration of the water snowline due to the infalling material during the disk formation could extend the region of planetesimal formation. More recently, \citet{Guilera+2020} and \citet{Lau+2022A&A} showed that static pressure bumps are also the preferential locations for the local formation of planetesimals due to the efficient dust accumulation at such locations. Moreover, \citet{Lau+2022A&A} showed that the gravitational stirring between planetesimals and the embryos that could form at the pressure bump can disperse planetesimals to broader regions. Finally, the works of \citet{ShibaikeAlibert2020A&A} and \citet{ShibaikeAlibert2023planetesimal} should also be mentioned, in which planetesimal formation happens in a pressure maximum developed at the outer edge of a migrating planet in a wide range of the protoplanetary disc. In this work, we present a new mechanism that triggers the formation of planetesimals at large distances from the central star and in broad regions of a few tens of au. 

Our paper is organized as follows: First, we briefly describe our code, whose detailed description is given in the Appendices, and then the particular disc model we use. Our results are presented in the third section: first with a modest jump in the $\alpha$ viscosity parameter that results in only a dust trap due to a traffic jam. The effect of larger jumps in the $\alpha$ viscosity parameter is also discussed. These large jumps result in real pressure maxima, in which solids are trapped for a long time while, due to the disc's viscous evolution, the pressure maxima exist. In the following subsections, the formation of a massive planetesimal is assumed also for the modest and stronger jump in viscosity that grows due to pebble accretion and later on by gas accretion. The paper closes with a summary and discussion of the results.

\section{Our numerical code and disc model}
To study the various effects that work at a transient density/pressure maximum, a numerical model has been developed that incorporates the following processes:
\begin{itemize}
    \item Gas evolution code in the vertically integrated and axis-symmetric case \citep{Lynden-BellPringle1974MNRAS}.
    \item N-body code using the Bulirsch-Stoer scheme incorporating the mutual gravitational interactions, as well as the planet disc interactions to follow the motion of the initially Moon-mass embryos \citep{Sandor+2011ApJ}, whose masses grow by pebble accretion \citep{Venturini+2020A&Aa, Venturini+2020A&Ab}.
    \item Partial gap opening in the gas surface density by the growing planets \citep{Chambers2021ApJ} that altering the gas surface density profile has a non-negligible impact on dust transport and evolution.
    \item The two-population model for dust transport and evolution \citep{Birnstiel+2012A&A} in the time-evolving disc, including sink terms due to pebble accretion by the growing solid planetary core and planetesimal formation via the streaming instability (SI), see the next point.
    \item Parametrized planetesimal formation via SI adopting various values of the planetesimal formation efficiency $\zeta$ \citep{Drazkowska+2016A&A}.
    \item Onset of giant planet formation and the collapse of the gaseous envelope to form a giant planet \citep{Ikoma+2000ApJ}. 
\end{itemize}
The equations describing the above phenomena are detailed in the Appendices. We note that in this particular work, only the formation, growth, and dynamical evolution of a single planet are investigated. 

\begin{figure}
   \centering
    \includegraphics[width=\columnwidth]{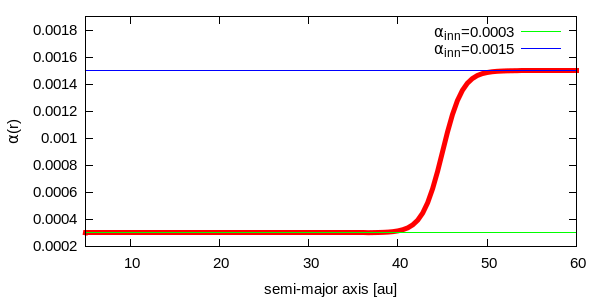}
      \caption{A viscosity jump when $\alpha_\mathrm{in}=0.0003$, $\delta\alpha=0.2$, $r_\mathrm{out}=45$ au, and $\Delta r_\mathrm{out}=2.25$ au.}
         \label{fig1:alpha_vs_r}
\end{figure}

A usual way to mimic the transitions between the regions of different turbulence is to apply an $\alpha$ viscosity parameter that is given as the function of the distance to the star:
\begin{equation}
\alpha(r) = \alpha_\mathrm{out} - \frac{\alpha_\mathrm{out}-\alpha_\mathrm{in}}{2}\left(1-\tanh \frac{r-r_\mathrm{out}}{\Delta r_\mathrm{out}} \right),
\label{eq:visc_reduction}
\end{equation}
being a smoothed step-function, where $r_\mathrm{out}$ is the radial location where the jump in viscosity occurs, see Figure \ref{fig1:alpha_vs_r}. The width of the transition region between the two regimes characterized by $\alpha_\mathrm{out}$ (for $r>r_\mathrm{out}$) and $\alpha_\mathrm{in}$ (for $r<r_\mathrm{out}$) is denoted by $\Delta r_{\mathrm{out}}$. To characterize the magnitude of viscosity transition we introduce 
\begin{equation}
\delta \alpha = \frac{\alpha_\mathrm{in}}{\alpha_\mathrm{out}}. 
\end{equation}
If $\delta\alpha=1$, there is no transition in viscosity, while the decreasing value of $\delta\alpha$ means a gradually stronger transition. 

The initial condition of the gas evolution equation is
\begin{equation}
  \Sigma_\text{gas,0}(r) = \frac{M_\text{disc}}{2 \pi R_c^2} \left( \frac{r}{R_c} \right)^{-1} e^{-(r/R_c)}, 
  \label{eq:initgasprofile}
\end{equation}
being parametrized with the disc mass $M_\text{disc}$ and critical radius $R_c$ for the disc compactness, with a viscosity prescription given by Equation \eqref{eq:visc_reduction}. Additionally, we use the following (fixed) temperature profile throughout our simulations
\begin{equation}
T(r) = 280 \left(\frac{r}{\mathrm{au}}\right)^{-0.5} \mathrm{K},
\end{equation}
assuming locally isothermal equation of state.

\begin{figure}[!h]
   \centering
    \includegraphics[width=\columnwidth]{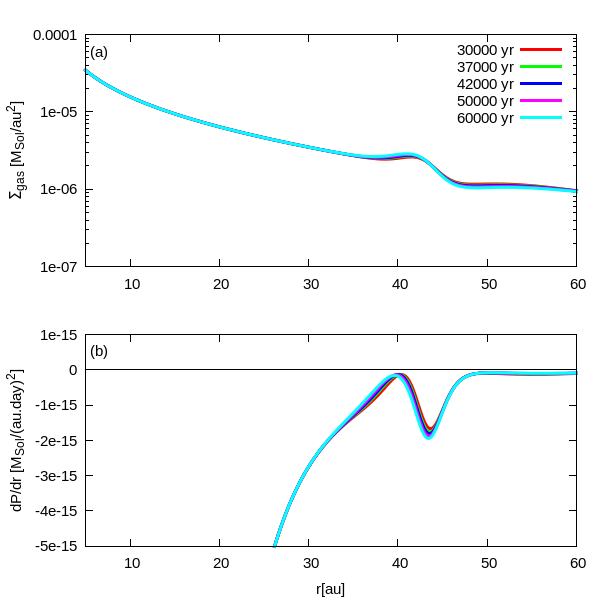}
    \includegraphics[width=\columnwidth]{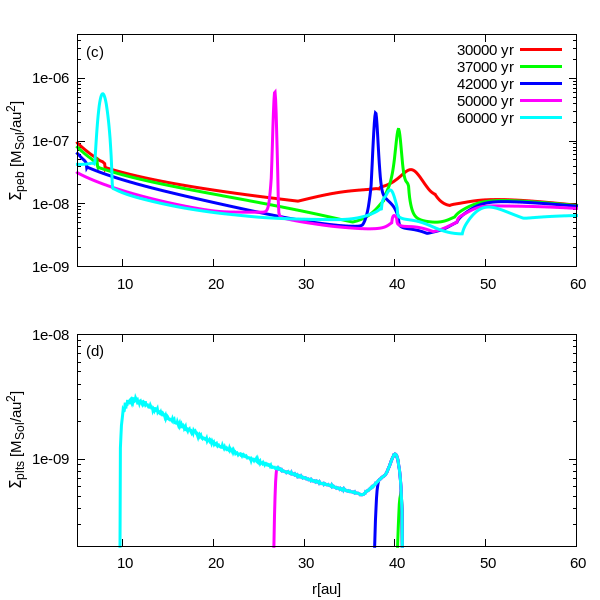}
      \caption{Disc evolution with $\delta\alpha = 0.2$ until the end of planetesimal formation. In panel (a) the $\Sigma_\mathrm{gas}(r)$ profiles are shown for different epochs. In panels (b), (c), and (d) the $\mathrm{d} P(r)/ \mathrm{d}r$, $\Sigma_\mathrm{peb}(r)$, and $\Sigma_\mathrm{plts}(r)$ profiles are shown at the same epochs as in panel (a).}
         \label{fig2:discevol_da0.2_nopl}
\end{figure}
\begin{figure}
   \centering
    \includegraphics[width=\columnwidth]{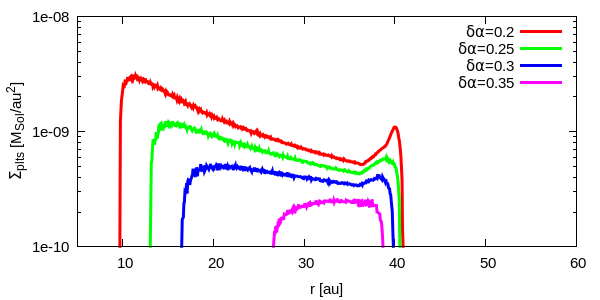}
      \caption{Surface density of planetesimals and the radial extent of their formation sites for different values $\delta\alpha$ of viscosity transition. The larger the value of $\delta\alpha$ the narrower the radial region in which planetesimals form.}
         \label{fig3:discevol_var_da_nopl}
\end{figure}

The initial condition for the dust transport equation is derived from Equation \eqref{eq:initgasprofile}:
\begin{equation}
   \Sigma_\text{dust,0}(r) = \epsilon \Sigma_\text{gas,0}(r),
\end{equation}
where $\epsilon$ is the solid-to-gas mass ratio, sometimes also referred to as disc metallicity.

In our simulations the following parameters are kept fixed: the initial dust-to-gas ratio $\epsilon=0.01$, initial size and density of dust grains $s_0 = 1\mu \mathrm{m}$, $\rho_\mathrm{dust}=1.6 \ \mathrm{g/cm}^3$. For the fragmentation threshold velocity of dusty ice particles, we use $u_\mathrm{frag} = 10 \ \mathrm{m/s}$, which has historically been used in the literature \citep{GundlachBlum2015ApJ}. The sticking and fragmenting properties of the ice-coated dust aggregates are, however, extremely important to resolve the so-called bouncing barrier \citep{Zsom+2010A&A}, therefore this field is the subject of intensive research. Laboratory experiments of \citet{MusiolikWurm2019ApJ} suggested that the water ice-coated dust particles behave similarly to silicates, seeming to invalidate the use of the value $10 \ \mathrm{m/s}$. More recently, \cite{Musiolik2021MNRAS} found that the UV irradiated ices of various volatile molecules on the surface of dust aggregates behave like liquids, therefore increasing their sticking properties and enabling higher velocity collisions that do not lead to the fragmentation of ice-coated dust aggregates.

We apply a planetesimal formation efficiency in the interval $\zeta\in[10^{-6}, 10^{-3}]$. In the main part of the paper those results are presented that utilize the highest value, $\zeta = 10^{-3}$, the same one proposed by \citet{DrazkowskaAlibert2017A&A}. However, \citet{Krapp+2019} have shown that the efficiency of planetesimal formation might be lower if the size distribution of pebbles is also considered. In a more recent paper, while applying a similar prescription for planetesimal formation, \citet{Izidoro+2022NatAs} used formation efficiency in the interval $\zeta\in[10^{-6}, 10^{-4}]$. 

In the MRI dead region, we use the viscosity parameter $\alpha_\mathrm{in}=0.0003$. This choice is motivated that even in the low ionized disc regions the vertical shear instability \citep{StollandKley2014A&A} generates a turbulent transport of gas at a similar level. In our simulations, we kept fixed this value, while we changed the value of $\alpha$ outside the MRI dead region in the interval $[7.5\times 10^{-4}, 6\times 10^{-3}]$ to control the strength of the viscosity jump, whose magnitude is an essential parameter of our study. These values, ranging between $3\times10^{-4}$ to $3\times10^{-3}$ or higher, are consistent with those estimated to reproduce the observed accretion rates from recent surveys, see \citet{Ribas2020A&A...642A.171R} and \citet{Rosotti2023} and references therein.
Additionally, the disc's mass is $M_\text{disc}=0.06 M_\odot$, the critical radius is $R_c=50$ au, and the viscosity transition happens at $r_\mathrm{out}=45$ au having a width $\Delta r_\mathrm{out}=2.25$ au, corresponding to the disc's scale height. The disc extends between 1 and 300 au divided by 5000 gridpoints placed equidistantly.

\begin{figure}[!h]
   \centering
    \includegraphics[width=\columnwidth]{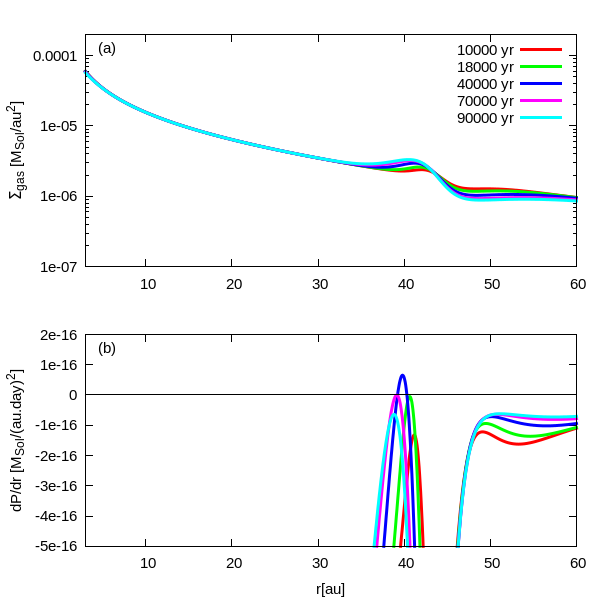}
    \includegraphics[width=\columnwidth]{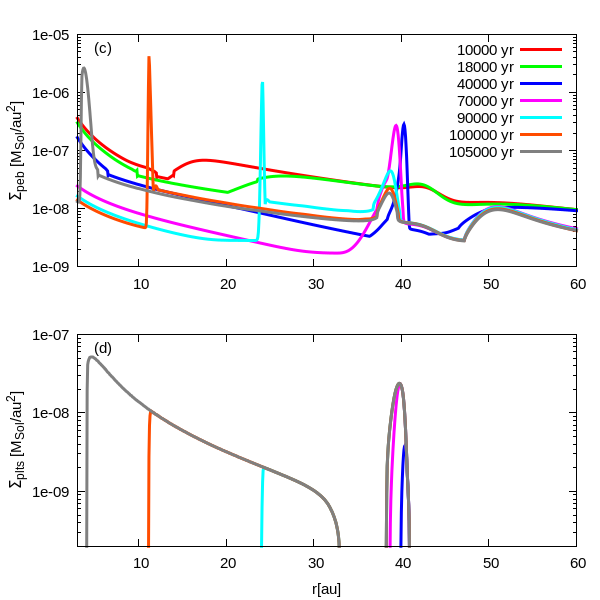}
      \caption{The same as in Figure \ref{fig2:discevol_da0.2_nopl} but with $\delta\alpha = 0.15$. In this case, a temporary pressure maximum develops. The $\Sigma_\mathrm{gas}(r)$ profiles are shown for different epochs in panel (a). In panel (b), the curves of the pressure gradient $\mathrm{d} P/ \mathrm{d}r$ are displayed at the same epochs as in the upper panel. Some of the curves of the pressure gradient cross the horizontal axis at zero, thus a temporary pressure trap forms between the epochs $1.8\cdot 10^4 < t < 7\cdot 10^4$ years. }
         \label{fig4:discevol_da0.15_nopl}
\end{figure}

\section{Results}
\subsection{Dust-trap due to a ``traffic jam''}
\label{subsec:traffic_jam}

In this part we present the outcome of our simulations when the changes in $\alpha$ viscosity are quite modest, falling in the interval:
\begin{equation}
0.2 \leq \delta \alpha  \leq 0.4.
\end{equation}
The following values are considered in our simulations: $\delta\alpha = 0.2$, 0.25, 0.3, and 0.35. In the first simulation $\delta \alpha = 0.2$, thus $\alpha_\mathrm{out}=0.0015$. In this case, as the disc evolves, a temporary maximum in $\Sigma_\mathrm{gas}$ develops (see panel (a) of Figure \ref{fig2:discevol_da0.2_nopl}), however, a corresponding maximum in $P(r)$ does not form as its derivative remains below zero $\mathrm{d} P/ \mathrm{d} r < 0$. (We recall that the gas pressure $P$ is given by Equation \eqref{eq:pressure}). On the other hand, the value of $\mathrm{d} P/ \mathrm{d} r$ is very close to zero (see panel (b) of Figure \ref{fig2:discevol_da0.2_nopl}) around the viscosity transition, therefore, the inward drift of solids is considerably slowed down, and as a consequence, dust particles are accumulated there and a ``traffic jam'' develops, as shown in panel (c). At a certain point, dust particles grow up to pebble size, e.g. their Stokes numbers become larger than our assumed threshold $\mathrm{St} \geq 0.01$ for streaming instability to be triggered. When the condition given by Equation \eqref{eq:streaming_cond} is also fulfilled, planetesimals begin to form according to Equation \eqref{eq:planetesimalformationeficiency} assuming $\zeta = 10^{-3}$ efficiency. Since $|\mathrm{d} P/ \mathrm{d} r|$ grows as $r\to 0$, the inward drift of pebbles becomes gradually faster, and planetesimals form in a wide range of the protoplanetary disc as the ring of concentrated pebbles rapidly drifts inward. This phenomenon is shown in panel (d) of Figure \ref{fig2:discevol_da0.2_nopl}.

To constrain the viscosity transition that still results in the formation of a visible amount of planetesimals, we run a few additional simulations with fixed $\alpha_\mathrm{inn} = 0.0003$ and slightly increasing $\delta \alpha = 0.25$, 0.3, 0.35, and 0.4 values, meaning gradually decreasing differences in the turbulence between the neighbouring layers. We have found that some amount of planetesimals could still be formed for $\delta \alpha = 0.35$, but not for $\delta\alpha=0.4$. On the other hand, the radial region in which planetesimals form is shrinking with increasing $\delta\alpha$. In Figure \ref{fig3:discevol_var_da_nopl} we show the surface density of planetesimals formed for the different values of $\delta\alpha$.

A final remark to our simulations is that even a short-living ($\sim 10^4$ year) transient dust-trap can trigger planetesimal formation in a broad region of the protoplanetary disc. In our case, the short-living dust trap is formed due to a traffic jam triggered by a modest jump in the $\alpha$ viscosity parameter. The existence of other mechanisms with similar outcomes cannot be, however, excluded \citep[e.g.][]{JiangOrmel2023MNRAS}. 

\subsection{Dust trap due to a pressure maximum}
\label{subsec:real_pressure_max}
To see the effect of a pressure maximum on the formation of planetesimals, we describe in this section the outcomes of two simulations, in which we use viscosity transitions $\delta\alpha = 0.15$ and $\delta\alpha = 0.05$.

In the simulation shown here, $\delta\alpha=0.15$ is close to the value at which no pressure maximum develops. We note, however, that the pressure maximum is transient in our presented simulations. The evolution of $\Sigma_\mathrm{gas}(r)$, $\mathrm{d}P(r)/\mathrm{d}r$, $\Sigma_\mathrm{peb}(r)$, and $\Sigma_\mathrm{plts}(r)$ profiles as the function of time is shown in panels (a), (b), (c), and (d) of Figure \ref{fig4:discevol_da0.15_nopl}. We see that during the disc viscous evolution, a pressure maximum develops at those radial locations where $\mathrm{d}P(r)/\mathrm{d}r=0$ fulfils and $\mathrm{d}P(r)/\mathrm{d}r$ is a strictly monotonically decreasing function of the distance $r$. The lifetime on the pressure maximum can thus be constrained between the epochs $1.8\cdot 10^4 < t < 7\cdot 10^4$ years, as seen in panel (b) of Figure \ref{fig4:discevol_da0.15_nopl}. During this time interval, pebbles accumulate at the maximum pressure, forming a narrow ring-like structure. In this ring of pebbles, the streaming instability is triggered as pebbles grow up to $\mathrm{St} > 0.01$ size and the condition given by Equation \eqref{eq:streaming_cond} is fulfilled. As the pressure maximum gradually vanishes the streaming instability is terminated; however, due to their inward radial drift, pebbles form another sharp peak (narrow ring), in which the streaming instability starts again. The rapidly inward drifting ring of pebbles creates planetesimals until reaching the semi-major axis at $~\sim 3$ au corresponding to the epoch $t\sim 10^5$ years. In this way, planetesimals form in a wide radial range of the disc quite rapidly, except an empty ring between $34\ \mathrm{au} < r < 37\ \mathrm{au}$, (see panel (d) of Figure \ref{fig4:discevol_da0.15_nopl}). When comparing the radial region of planetesimal formation, one can see that planetesimals are formed in a wider region than in our previous simulations with $\delta\alpha \geq 0.2$. The reason for this result is not very surprising, even a short-living pressure maximum can efficiently collect pebbles forming a massive ring that can survive for a longer time drifting closer to the star.
\begin{figure}[!h]
   \centering
    \includegraphics[width=\columnwidth]{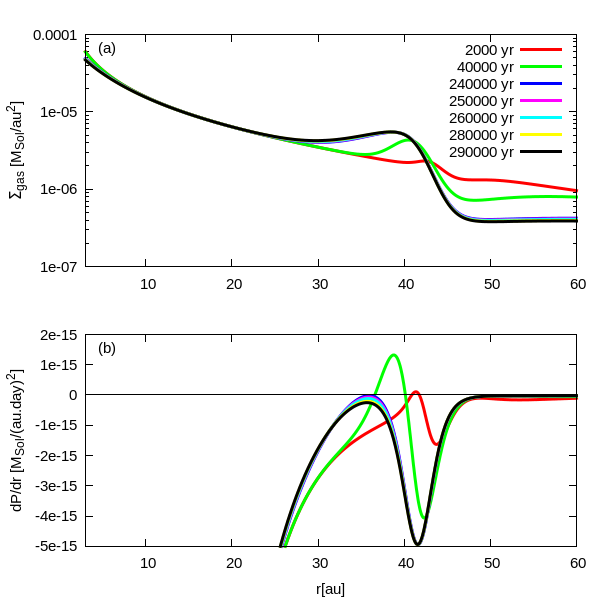}
    \includegraphics[width=\columnwidth]{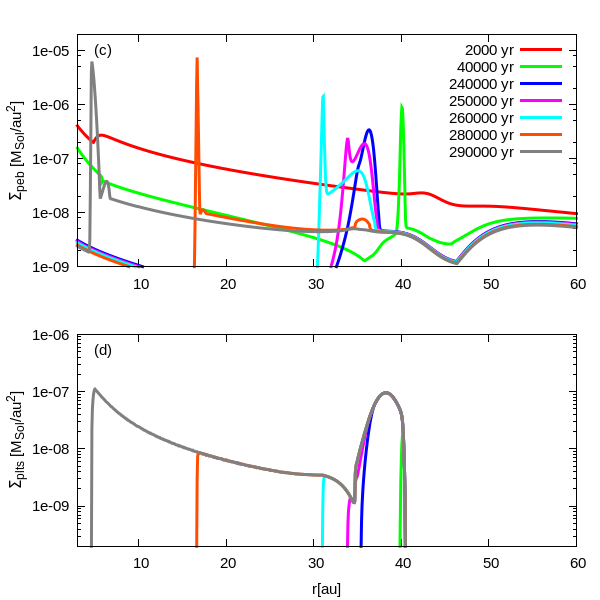}
      \caption{The same as in Figure \ref{fig4:discevol_da0.15_nopl} but with $\delta\alpha = 0.05$. In this case, a temporary pressure maximum develops, too. The $\Sigma_\mathrm{gas}(r)$ profiles are shown for different epochs in panel (a). In panel (b), the curves of the pressure gradient $\mathrm{d} P/ \mathrm{d}r$ are displayed at the same epochs as in the upper panel. Some of the curves of the pressure gradient cross the horizontal axis at zero, thus a temporary pressure trap forms between the epochs $2\cdot 10^3 < t < 2.4\cdot 10^5$ years.}
         \label{fig45:discevol_da0.05_nopl}
\end{figure}

To see the effect of the parameter $\delta\alpha$ on the formation of planetesimals, we redo the above simulation using $\delta\alpha = 0.05$ which represents a stronger jump in viscosity. The results of this simulation are displayed in Figure \ref{fig45:discevol_da0.05_nopl}. Similarly to the previous case with $\delta\alpha = 0.15$, a pressure maximum develops but survives for a much longer time $\sim 2\cdot 10^5$ years. During this time, pebbles are assembled in a ring-like structure around the pressure maximum and some amount of them are transformed into planetesimals. When the pressure maximum is terminated the ring of pebbles drifts inwards and planetesimals still form in a narrow ring, as the corresponding peaks show in panel (c) of Figure \ref{fig45:discevol_da0.05_nopl}. In panel (d), the surface density of planetesimals is shown at the epochs displayed in the panels. In this case, there is no gap in the surface density of planetesimals formed, and they form in a larger quantity. A larger jump in viscosity results in a larger overall mass of planetesimals. 

During this quite long lifetime of the pressure maximum, the formation of a larger body (planetesimal or planetary core) can also be assumed that can further grow by pebble accretion. In what follows, we describe our results if the formation of a more massive planetesimal or an embryo is considered in the ring of pebbles and planetesimals.

\subsection{Effects of a growing embryo}
\label{subsec:grembryo}
If a massive enough body (planetesimal or planetary embryo) can form in the concentrated ring of pebbles, it might be able to further grow by pebble accretion \citep{LambrechtsJohansen2012A&A}. It has been shown that the fast growth of such a body can lead to the formation of a giant planet, \citep[see][]{Guilera+2020}. In another study, \citet{Chambers2021ApJ} used bodies with initial mass $10^{-4} M_\oplus$ in his simulations to grow further via pebble accretion. In a more recent investigation, \citet{Lau+2022A&A} did self-consistent simulations on the formation of multiple planetary cores and planetesimals in a relatively wide region around the outer edge of a gap. They have found $\sim 3\cdot 10^{-3} M_\oplus$ as a characteristic planetesimal mass. Inspired by the above works, in our simulations, we insert in the pressure maximum one body with the following initial masses: (i) $10^{-2} M_\oplus$, (ii) $3\cdot 10^{-3} M_\oplus$, and (iii) $10^{-4} M_\oplus$, each of them corresponding to a separate simulation. Conservatively, we inserted this body when the overall mass of planetesimals exceeded $5 \cdot 10^{-2} M_\oplus$ at the pressure maximum. We are aware of the fact that probably not only one planetesimal is formed in the pressure trap \citep[see][]{Lau+2022A&A}, however, here we assume that there exists a body which is the more massive one, grows faster than the others. 

Since in our work pebble accretion is assumed to be the main mechanism for the growth of the large planetesimals or planetary cores, it is important to check whether the inserted body is in the Bondi (headwind) or Hill (shear) regime \citep{LambrechtsJohansen2012A&A,Ormel2017ASSL,Lyra+2023ApJ}. Usually, pebble accretion in the Hill regime is more efficient, however, as a recent study of \citet{Lyra+2023ApJ} showed, the 3D polydisperse (using all sizes from the size distribution of pebbles) Bondi accretion could also be quite efficient. Therefore we check the mass of the inserted body to know in which regime is. 

For pebble accretion in the Hill regime, the mass of the inserted body should be larger than the transition mass between the Bondi and Hill regimes ($M_\mathrm{HB}$), which is given by \cite{Ormel2017ASSL}:
\begin{equation}
    M_\mathrm{HB} = \frac{M_\mathrm{t}}{8\mathrm{St}},
    \label{eq:HillBondi_limit}
\end{equation}
where $\mathrm{St}$ is the Stokes number of the particle and $M_\mathrm{t}$ can be written as
\begin{equation}
    M_\mathrm{t} = \frac{\Delta v^3}{G\Omega_\mathrm{Kep}},
\end{equation}
being $G$ the constant of gravity, $\Omega_\mathrm{Kep}$ the angular velocity for the circular Keplerian orbit, and $\Delta v$ the relative velocity between pebbles and the growing body. This relative velocity can be easily calculated as
\begin{equation}
    \Delta v = \frac{\sqrt{4 \mathrm{St}^2 + 1}}{\mathrm{St}^2 + 1}|\eta| v_\mathrm{Kep},
\end{equation}
where $\eta$ is
\begin{equation}
    \eta = -\frac{h^2}{2}\frac{r}{P}\frac{\mathrm{d}P}{\mathrm{d}r},
\end{equation}
being $h$ the gas disc's aspect ratio and $P$ the gas pressure. 

\noindent
The limit for the 3D polydisperse Bondi accretion \citep{Lyra+2023ApJ}, which can also be very effective is
\begin{equation}
    M_\mathrm{3DBondi} = \frac{\Delta v\Omega_\mathrm{Kep}H_\mathrm{gas}\alpha}{G \mathrm{St}(\mathrm{St}+\alpha)}.
    \label{eq:3DBondi_limit}
\end{equation}

\begin{figure}
   \centering
    \includegraphics[width=\columnwidth]{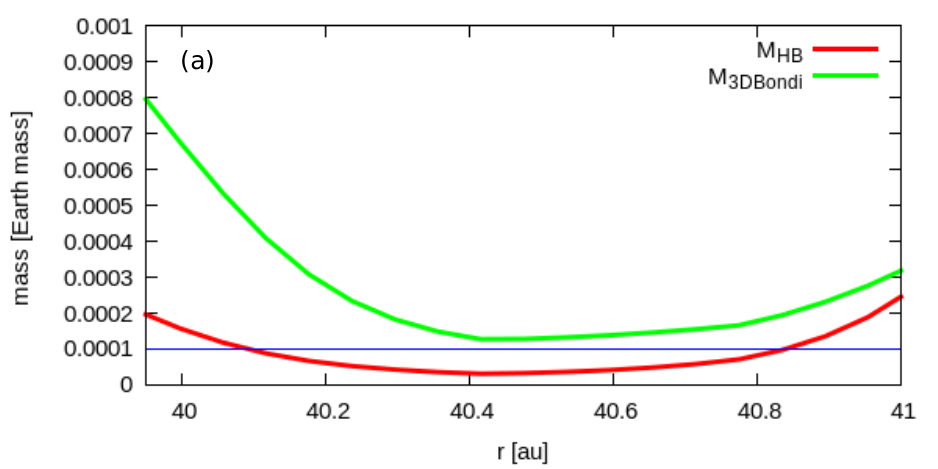}
    \includegraphics[width=\columnwidth]{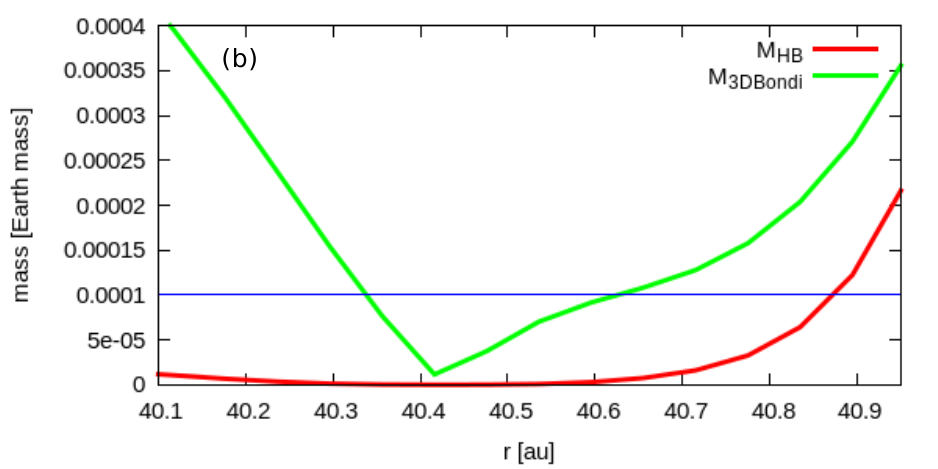}
    \includegraphics[width=\columnwidth]{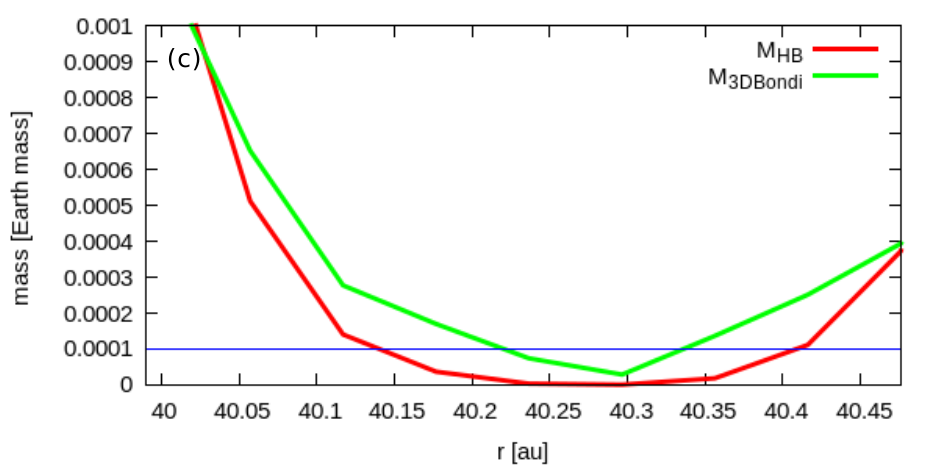}
    \caption{The transition masses ($M_\mathrm{HB}$ and $M_\mathrm{3DBondi}$) as the function of the radial distance around the dust traps. In panel (a) $\delta\alpha = 0.2$, when the accumulation of solids occurs due to a traffic jam. In panels (b) and (c), those cases are shown, in which pressure maxima are developed for the cases of $\delta\alpha=0.15$ and $\delta\alpha=0.05$, respectively. The horizontal line displays the minimum mass of the inserted embryo in our simulations ($10^{-4}M_\oplus$).}
         \label{fig:trmasses}
\end{figure}

Since the relative velocity $\Delta v$ tends to zero when approaching the radial position of the pressure maximum, the transition mass can also be arbitrarily small there. On the other hand, the radial position of the pressure maximum is slightly moving inwards, therefore it is important to know the most convenient radial position where the massive planetesimal or embryo should be inserted to be in the Hill regime of pebble accretion. To determine the optimal location of the embryo, we plot the transition masses as functions of the distance from the star in that region when planetesimal formation took place until the overall mass of planetesimals reached the $5 \cdot 10^{-2} M_\oplus$ value, see panels (a), (b), and (c) of Figure \ref{fig:trmasses}.

We can see that even in the case of the traffic jam, the inserted body with the mass $10^{-4}M_\oplus$ is in the Hill regime of pebble accretion, however, according to Equation \eqref{eq:3DBondi_limit}, it can also be subject of the 3D polydisperse Bondi accretion. In the other two cases, this mass is above $M_\mathrm{3DBondi}$, however in a relatively narrow region. The other two masses, $3\cdot 10^{-3}M_\oplus$ and $10^{-2}M_\oplus$ are reassuringly in the Hill regime of pebble accretion. In the main part of our paper, we present those simulations that correspond to this latter case, namely a Moon-mass core is inserted, when the overall mass of planetesimals became larger than $5 \cdot 10^{-2} M_\oplus$. The results of our simulations using smaller mass bodies are shown in the Appendix. 

As a final remark, we should note that the growth process leading to the formation of this core is not modelled in our simulations, however, based on earlier investigations, we assume that the formation of such a body might physically be reasonable by runaway growth of planetesimals \citep{Liu+2019A&A}, formation of large planetesimals \citep{Lau+2022A&A}, or by gravitational collapse of the ring of pebbles \citep{Takahashi+2023ApJ}.

\subsubsection{Formation of a planet in the ``traffic jam'' ($\delta\alpha=0.2$) by pebble accretion}
\begin{figure}
   \centering
    \includegraphics[width=\columnwidth]{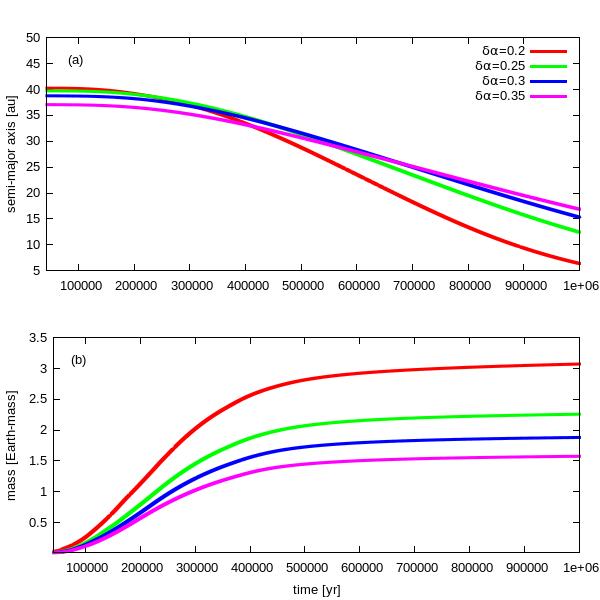}
      \caption{Inward migration (panel (a)) and mass growth of the planet due to pebble accretion (panel (b)) during the whole length of the numerical simulation for various values of $\delta\alpha$. The evolution of the mass and the semi-major axis of the planet for each value of $\delta\alpha$ is shown with different colours.}
         \label{fig5:migr_massgr_trjam}
\end{figure}

The massive planetary body corresponding to case (i) is inserted in the initially assembled ring of pebbles around $r\approx 40$ au.  The mass of this planetary core is growing further by pebble accretion, and when the maximum in $\Sigma_\mathrm{gas}$ is gradually diminishing and the planet trap is terminated, the planetary core begins to migrate towards the star. In this particular simulation with $\delta\alpha=0.2$ the core reaches the mass of a Super-Earth, $m_\mathrm{pl} = 3 M_\oplus$ until arriving at $a_p=7$ au during the whole numerical simulation being, in this case, $10^6$ (see Figure \ref{fig5:migr_massgr_trjam}). During the mass growth and inward migration, the planetary core does not reach the critical core mass given by Equation \eqref{eq:critical_coremass} that would lead to the formation of a giant planet. 

In Figure \ref{fig5:migr_massgr_trjam} we also show our results by using gradually growing viscosity jump parameters $\delta\alpha$. The larger the value of $\delta\alpha$ (meaning a more modest jump), the smaller the mass of the planet formed. We have found in our simulations that the masses of planets are between $1.5-3 M_\oplus$. As a consequence, the smaller mass planet does not migrate so close to the central star, as the more massive one. In the Appendix, we present how the mass and migration history of the planet formed depends on the initial mass of the inserted body into the pressure trap.

\begin{figure}[!t]
   \centering
    \includegraphics[width=\columnwidth]{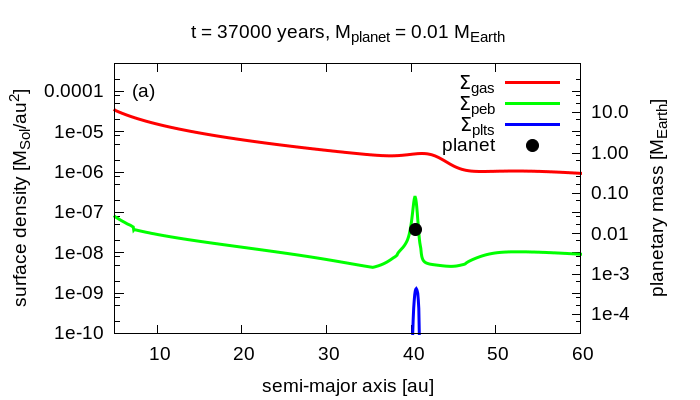}
    \includegraphics[width=\columnwidth]{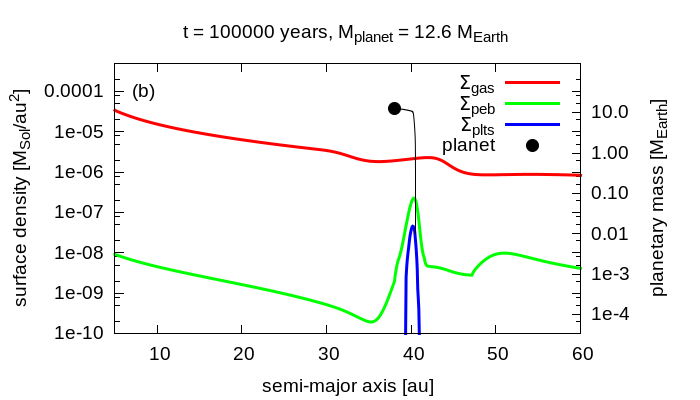}
    \includegraphics[width=\columnwidth]{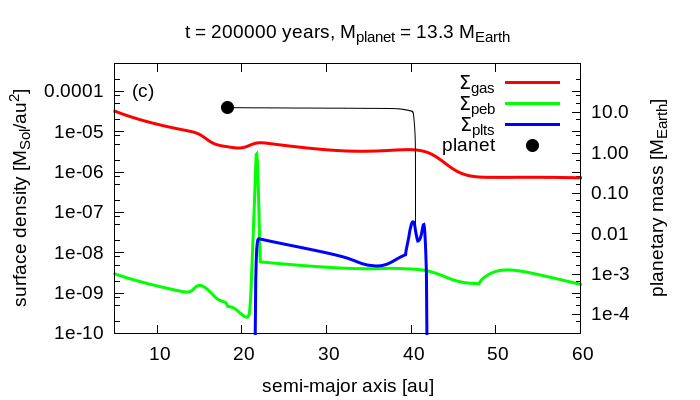}
    \includegraphics[width=\columnwidth]{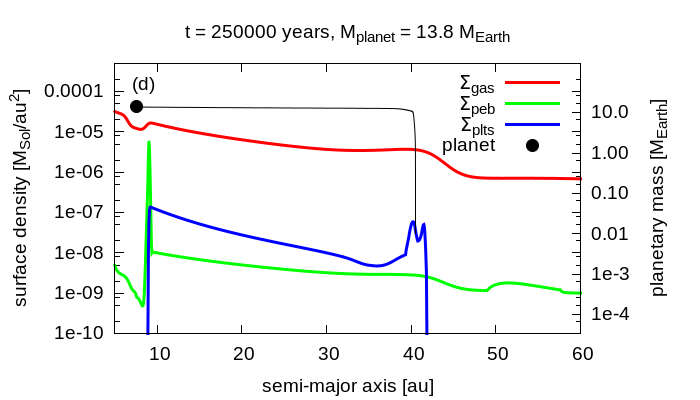}
      \caption{Episodes of the mass growth and inward migration of a planet that forms at the boundary of two regions with a viscosity transition $\delta\alpha = 0.15$. The evolution of $\Sigma_\mathrm{gas}(r)$, $\Sigma_\mathrm{peb}(r)$, and $\Sigma_\mathrm{plts}(r)$ are shown together with the migration and mass growth of the planet. The large black dot, and the black line indicate the planet's position and its track in the $a_\mathrm{pl}-m_\mathrm{pl}$ plane.}
         \label{fig6:sdens_etc_da0.15_mNep}
\end{figure}
\begin{figure}[]
   \centering
    \includegraphics[width=\columnwidth]{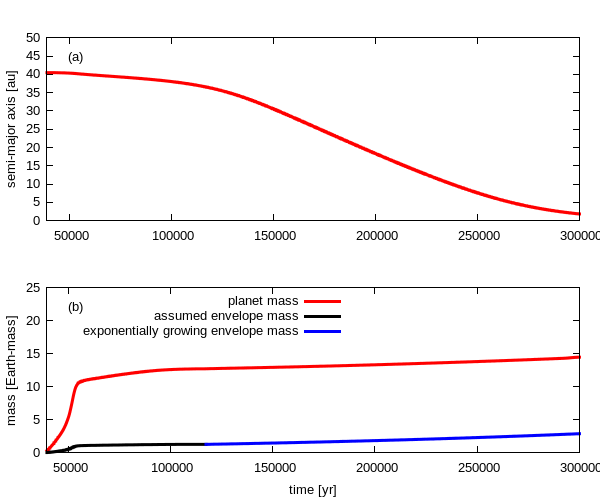}
      \caption{Inward migration (panel (a)) and mass growth of the planet due to pebble accretion and formation of a massive gaseous envelope (panel (b)) are shown in the case $\delta\alpha=0.15$. The black line in the bottom panel reflects the assumption that the envelope's mass is 10\% of the total mass until the exponential growth of the envelope is shown with the blue line.}
      \label{fig7:migr_mgr_da0.15_mNep}
\end{figure}
\subsubsection{Formation of a planet in the pressure maximum ($\delta\alpha=0.15$) by pebble accretion}
The various effects leading to the formation of planetesimals as well as a mini-Neptune mass planet are shown in the panels of Figure \ref{fig6:sdens_etc_da0.15_mNep}, where the red line shows the surface density of gas $\Sigma_\mathrm{gas}(r)$, the green line the surface density of solids $\Sigma_\mathrm{peb}(r)$ (mostly pebbles in our case), the blue line the surface density of planetesimals formed $\Sigma_\mathrm{plts}(r)$. The black dot denotes the planet and the black line is its evolutionary track on the semi-major axis -- planet mass plane. 

As our simulations show, right after the development of a density maximum in $\Sigma_\mathrm{gas}(r)$, a maximum in $P_\mathrm{gas}(r)$ develops, too. This pressure maximum traps dust particles very efficiently that grow rapidly to sizes above $\mathrm{St} \geq 0.01$, and when the condition given by Equation \eqref{eq:streaming_cond} fulfils, streaming instability is triggered and planetesimals begin to form. Around the epoch $t\sim 37000$ years, displayed in panel (a), when the overall mass of planetesimals formed exceeds 5 Moon-mass, we insert a Moon-mass planetary core that is subject to grow further by pebble accretion, see Equations \eqref{eq:pebble_accr_rate2D} and \eqref{eq:pebble_accr_rate3D}. The initial Moon-mass core grows rapidly and at $t\sim 10^5$ years its mass is about 12.6 $M_\oplus$, see panel (b). Until this epoch, the planet has already migrated a bit inward from its place of formation following the disc's viscous evolution and according to Equation \eqref{eq:gapopening}, a partial gap in $\Sigma_\mathrm{gas}(r)$ has already been opened. The peak in the surface density of pebbles $\Sigma_\mathrm{peb}(r)$ is trapped at the outer edge of this gap and moves together with the planet. At the epoch $t\sim 2\cdot 10^5$ years, the gap is deepened and widened as the planet's mass grows up to 13.3 $M_\oplus$, as shown in panel (c).  
Finally, at the epoch $t\sim 2.5 \cdot 10^5$ years, shown in panel (d), the planet's mass is $m_\mathrm{pl}\sim 13.8 M_\oplus$, and it has migrated to $a_\mathrm{pl}\sim 7.6$ au. As the partial gap migrates with the planet, the peak in $\Sigma_\mathrm{peb}(r)$ follows the planet, and since the conditions of the streaming instability are fulfilled, planetesimals still form, as can be seen in $\Sigma_\mathrm{plts}(r)$ curve. 

The mass growth and the migration history of the planet are shown in panels (a) and (b) of Figure \ref{fig7:migr_mgr_da0.15_mNep}, respectively. Around the epoch $t\sim 1.17\cdot 10^5$ years the flux of pebbles is drastically reduced thus the planet reaches the critical mass given by Equation \eqref{eq:critical_coremass}, as no heating mechanism can support the gaseous envelope in hydrostatic equilibrium. This sudden reduction of the flux of pebbles also means that the planet reaches the pebble isolation mass \citep{Lambrechts+2014}. In our work, pebble isolation is obtained self-consistently, only the gap opened by the planet regulates the flux of pebbles onto the planet. From this epoch, the mass of the planet grows solely due to gas accretion by the already existing envelope, which grows exponentially on a Kelvin-Helmholtz timescale, see Equations \eqref{eq:envelope_growth} and \eqref{eq:KH_time}. Instead of solving the equations for the structure and evolution of the initial gaseous envelope, $M_\mathrm{env}^0$ is assumed to be 10 \% of the total planet mass \citep{Schneider+2021A&A} being formed mainly by evaporated pebbles, as the black curve in panel (b) of Figure \ref{fig7:migr_mgr_da0.15_mNep} shows. Although the mass growth of the envelope is exponential, during the planet's migration there is not enough time for the formation of a giant planet. Since the total mass of the planet at the end of its migration does not exceed $20 M_\oplus$ we assumed type I migration throughout the simulation. While the planet has been acquiring the mass of a mini Neptune-like planet, at the outer edge of the gap planetesimals have still been forming during its inward migration, as seen in panel (b) of Figure \ref{fig6:sdens_etc_da0.15_mNep}.   

\subsubsection{Formation of a planet in the pressure maximum ($\delta\alpha=0.05$) by pebble accretion}
\begin{figure}[]
   \centering
    \includegraphics[width=\columnwidth]{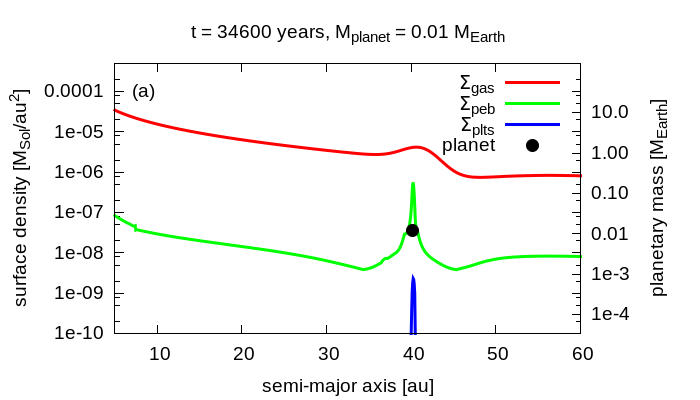}
    \includegraphics[width=\columnwidth]{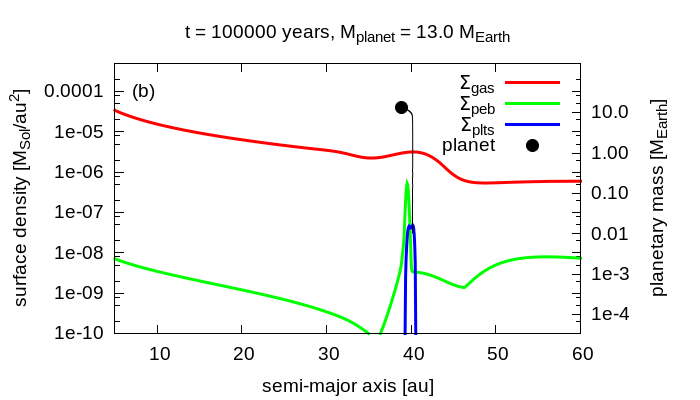}
    \includegraphics[width=\columnwidth]{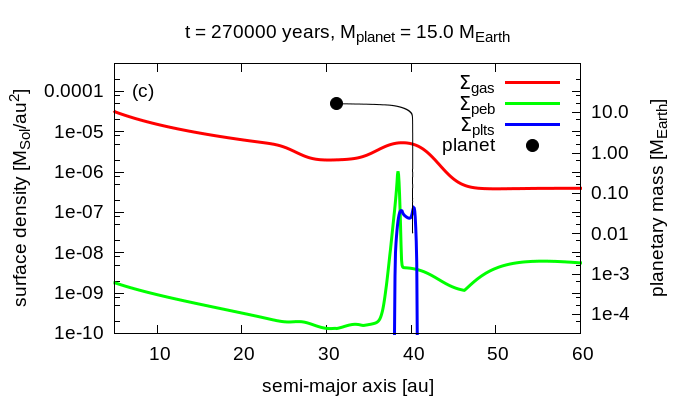}
    \includegraphics[width=\columnwidth]{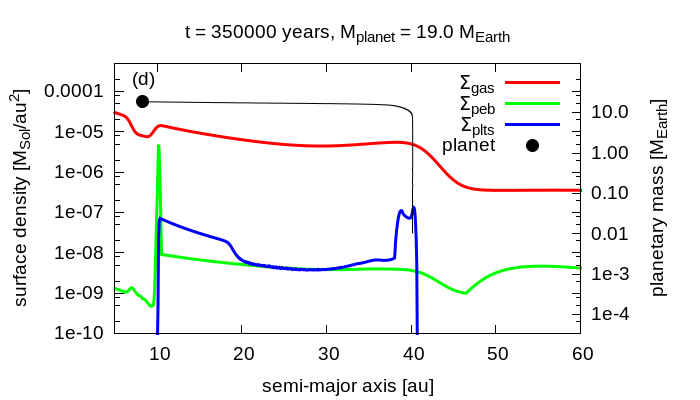}
      \caption{The same as in Figure \ref{fig6:sdens_etc_da0.15_mNep} with a viscosity transition $\delta\alpha = 0.05$.}
         \label{fig8:sdens_etc_da0.05_mNep}
\end{figure}

\begin{figure}[]
   \centering
    \includegraphics[width=\columnwidth]{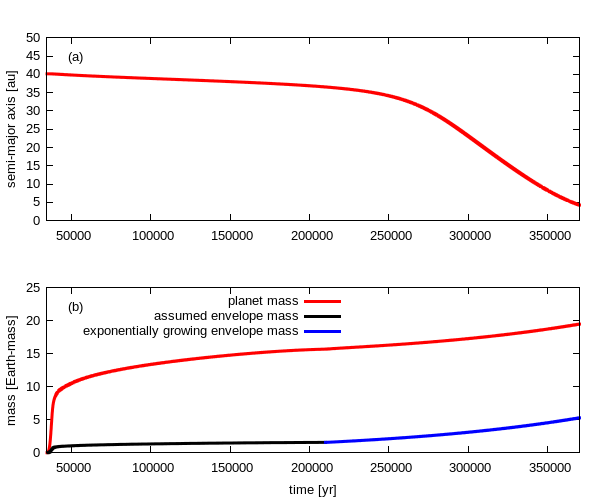}
      \caption{The same as in Figure \ref{fig7:migr_mgr_da0.15_mNep} in the case of $\delta\alpha=0.05$.}
      \label{fig9:migr_mgr_da0.15_mNep}
\end{figure}

To investigate how the value of $\delta\alpha$ affects the outcome of the above-described scenario, we have also done a simulation with $\delta\alpha=0.05$. In this case, the various formation episodes of planetesimals and Neptune-mass planets (with final mass $\sim 20 M_\oplus$) are displayed in the panels of Figure \ref{fig8:sdens_etc_da0.05_mNep}. The overall scenario is however very similar to the case of $\delta\alpha=0.15$. By comparing Figure \ref{fig7:migr_mgr_da0.15_mNep} to Figure \ref{fig9:migr_mgr_da0.15_mNep} it can be seen that in the simulation with $\delta\alpha=0.15$ a Neptune mass planet forms with mass $\sim 15 M_\oplus$. This planet, however, migrates closer to the star in a shorter time. This behaviour can be explained by the more modest jump in the viscosity, which results in the termination of the planet trap in a shorter time than in the simulation with $\delta\alpha=0.05$. Being trapped for a shorter time in the pebble-rich environment, the planet acquires a lower mass. Due to the earlier release from the planet trap, the planet also gets closer to the star. Moreover, since the accretion rate of the gaseous envelope depends on the core's mass when the pebble accretion practically stops, the gaseous envelope is also less massive than in the case of $\delta\alpha=0.05$. These simulations support the conjecture that the smaller the value of $\delta\alpha=0.05$ (e.g. the stronger the jump of viscosities), the larger the mass of the formed planet.

Contrary to the previous simulations with $\delta\alpha \geq 0.2$ presented in Subsection \ref{subsec:traffic_jam}, the solid core forms first, which is followed by the formation of planetesimals in a wide radial range of the disc at later epochs. In this sense, planetesimals are byproducts of the formation of a (mini) Neptune-like planet during its inward migration. We note that planetesimal formation at the outer edge of the gas gap opened by a migrating planet has already been proposed by \citet{ShibaikeAlibert2020A&A} and \citet{ShibaikeAlibert2023planetesimal}, however, in our work the formation and growth of the planet are modelled in a more comprehensive physical model.

\section{Effects of the planetesimal formation efficiency and the migration speed of the planet}
In this part, we show how the final surface density of planetesimals and the mass of the formed planet depend on the planetesimal formation efficiency $\zeta$ and the migration speed of the planet.

\subsection{The amount of planetesimals formed without planet formation}
Similar to the 3.1 and 3.2 subsections, here we present the results of our simulations for lower values of the planetesimal formation efficiency $\zeta$. We run several simulations for $\zeta = 10^{-4}$, $10^{-5}$, and $10^{-6}$ in cases of a traffic jam ($\delta\alpha=0.2$) and dust traps due to pressure maxima ($\delta\alpha=0.15$, $0.05$). We note that in these cases we do not enable the formation of a large planetesimal/small planetary core from the ring of pebbles accumulated at the viscosity transition being the subject of further growth by pebble accretion.

\begin{figure}
   \centering
    \includegraphics[width=\columnwidth]{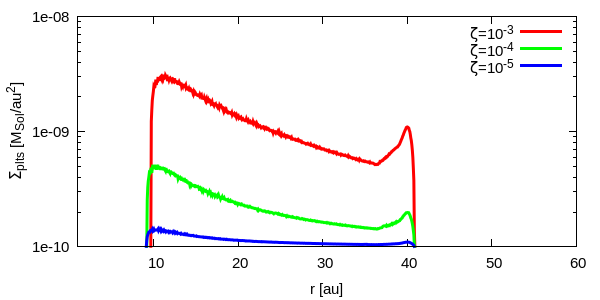}
      \caption{Surface density of planetesimals for different values of planetesimal formation efficiency $\zeta=10^{-3}$, $10^{-4}$, and $10^{-5}$ when $\delta\alpha=0.2$. The larger the value of $\zeta$ the larger the overall mass of planetesimals formed.}
         \label{fig12:planetesimals_low_zeta_traffic_jam_no_planet}
\end{figure}
\begin{figure}
   \centering
    \includegraphics[width=\columnwidth]{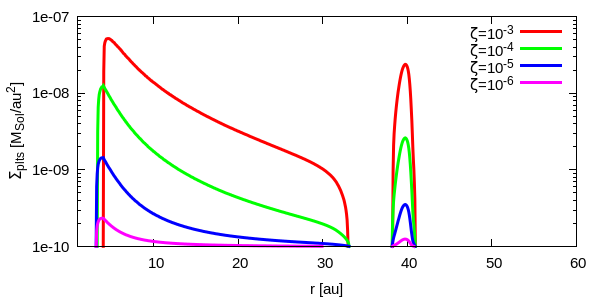}
      \caption{Surface density of planetesimals for different values of planetesimal formation efficiency $\zeta=10^{-3}$, $10^{-4}$, $10^{-5}$, and $10^{-6}$ when $\delta\alpha=0.15$. The larger the value of $\zeta$ the larger the mass of planetesimals formed.}
         \label{fig13:planetesimals_low_zeta_p_max_no_planet}
\end{figure}
\begin{figure}
   \centering
    \includegraphics[width=\columnwidth]{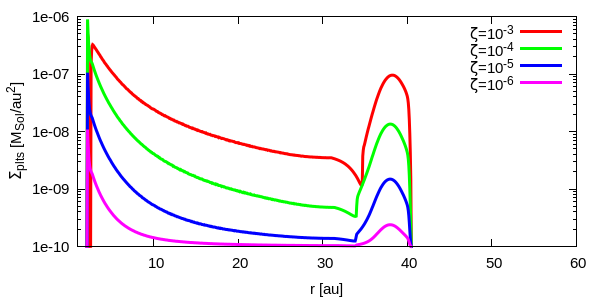}
      \caption{Surface density of planetesimals for different values of planetesimal formation efficiency $\zeta=10^{-3}$, $10^{-4}$, $10^{-5}$, and $10^{-6}$ when $\delta\alpha=0.05$. The larger the value of $\zeta$ the larger the mass of planetesimals formed.}
         \label{fig14:planetesimals_low_zeta_p_max_no_planet}
\end{figure}
In Figure \ref{fig12:planetesimals_low_zeta_traffic_jam_no_planet}, we present the case of $\delta\alpha=0.2$ with $\zeta \in [10^{-5}, 10^{-3}]$. We note that in this case no pressure maximum develops in gas, and pebbles are accumulated due to a traffic jam. According to the expectations, the overall mass of planetesimals is decreasing with lower values of $\zeta$. On the other hand, planetesimal formation is triggered, and happens in the same radial extent of the disc, since the condition given by Eq. \eqref{eq:streaming_cond} does not depend on $\zeta$. 

In Figures \ref{fig13:planetesimals_low_zeta_p_max_no_planet} and \ref{fig14:planetesimals_low_zeta_p_max_no_planet} the surface density of planetesimals are shown for those cases with $\delta\alpha = 0.15$ and $0.05$, in which a real pressure maximum develops at the viscosity transition. We note that the orange colour lines correspond to cases of $\zeta = 10^{-3}$, already shown in Figures \ref{fig4:discevol_da0.15_nopl} and \ref{fig45:discevol_da0.05_nopl}. From these figures, one can immediately see that lower $\zeta$ values result in less amount of planetesimals. 

\subsection{The amount of planetesimals and the mass of the formed planet}
In this part, the amount of planetesimals and the mass of the planet are investigated for gradually decreasing values of the parameter $\zeta$. Contrary to the case presented in the previous subsection, here we insert a massive planetesimal (or small mass planetary core) to the pressure maximum when the overall mass of planetesimals formed around the pressure maximum exceeds the mass of the Moon ($10^{-2} M_\oplus$). We use two values for the mass of the inserted body, namely $10^{-3} M_\oplus$ and $10^{-4} M_\oplus$ being one tenth and one hundredth of the Moon mass, respectively. The mass of the inserted body does not affect the outcome of our simulations.

\begin{figure}
   \centering
    \includegraphics[width=\columnwidth]{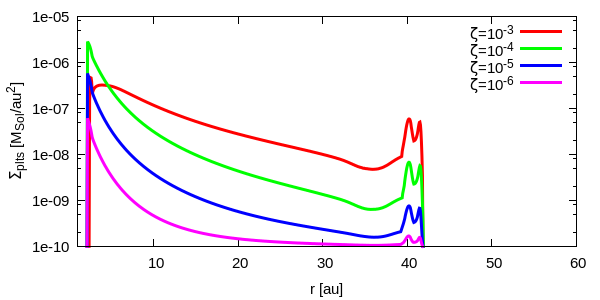}
      \caption{Surface density of planetesimals for different values of planetesimal formation efficiency $\zeta=10^{-3}$, $10^{-4}$, $10^{-5}$, and $10^{-6}$ when a massive planetesimal is inserted in the pressure maximum, which develops as the result of the viscosity jump $\delta\alpha=0.15$. The larger the value of $\zeta$ the larger the overall mass of planetesimals formed.}
         \label{fig15:planetesimals_low_zeta_p_max_w_planet}
\end{figure}

In Figure \ref{fig15:planetesimals_low_zeta_p_max_w_planet} the surface density of planetesimals formed is shown when using various planetesimal formation efficiencies. The pressure maximum develops due to the viscosity jump $\delta\alpha = 0.15$, in which a massive planetesimal is inserted. The smaller the parameter $\zeta$ the smaller the overall mass of planetesimals formed. We have also examined how the change in $\zeta$ affects the final mass of the planet formed. Interestingly this value is not particularly sensitive on $\zeta$: for $\zeta = 10^{-3}$ the mass is 14.86 $M_\oplus$, in the interval $\zeta \in[10^{-6}, 10^{-4}]$ the mass of the planet ranges between 15.24 and 15.46 $M_\oplus$. The explanation for this slight change could be that due to the reduced planetesimal formation efficiency, the amount of pebbles for small $\zeta$ is larger in the pressure maximum, therefore the pebble accretion rate is also slightly higher, leading to higher solid core mass.

When the massive planetesimal is inserted to the pressure maximum that forms due to the viscosity jump $\delta\alpha=0.05$, the mass of the planet formed ranges between 21.26 and 21.56 $M_\oplus$ for $\zeta \in[10^{-6}, 10^{-3}]$, while the surface density of planetesimals changes very similarly to the case shown in Figure \ref{fig15:planetesimals_low_zeta_p_max_w_planet}, therefore we do not display here.

Our simulations show that according to expectations the reduction of the planetesimal accretion efficiency leads to less overall mass of planetesimals formed, while in those cases when planet formation is also assumed and investigated, the final mass of the planet formed is practically not sensitive to $\zeta$. 

\subsection{The effect of the migration speed on the mass of planetesimals and the mass of the planet formed}
In our simulations due to the present capability of our model, we use a locally isothermal equation of state, meaning that the radial profile of the disc temperature is kept stationary. To keep our model simple, we, therefore, implemented the prescription of \citet{Tanaka+2002ApJ} for migration and that of \citet{TanakaWard2004ApJ} for damping the planet's eccentricity and inclination. This prescription, however, may result in too fast inward migration, so instead of considering more elaborated ones in the present study, we have investigated a few cases, in which the migration timescale is artificially increased. In this way, we can study the effects of reduced migration rates on the formation of planetesimals and the mass of the planet.

\begin{figure}[]
   \centering
    \includegraphics[width=\columnwidth]{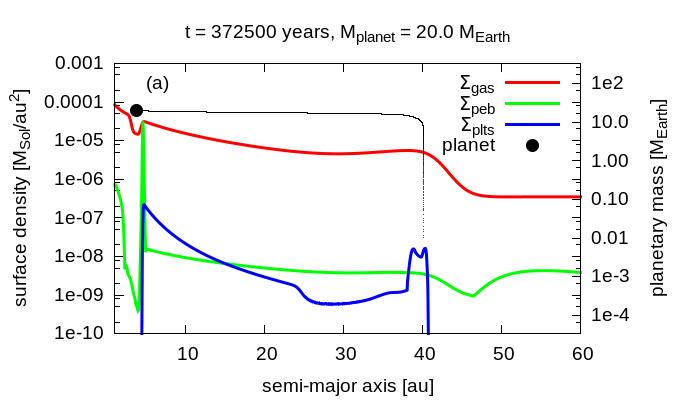}
    \includegraphics[width=\columnwidth]{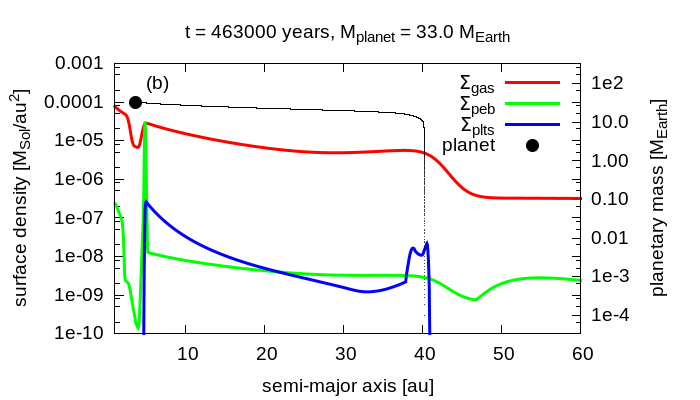}
    \includegraphics[width=\columnwidth]{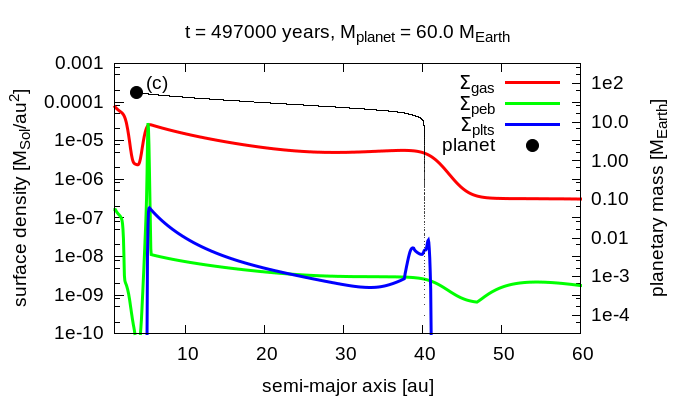}
      \caption{The results of our simulations using parameters $\delta\alpha = 0.05$, $\zeta = 10^{-4}$, and reduced migration speed for planet migration. In panel (a) the reference run is displayed with unmodified migration speed. In panel (b) the migration timescale is twice, while in panel (c) is three times when compared to the migration time displayed in panel (a). All simulations are done until the planet reaches 3.7 au (which is chosen arbitrarily). We recall that on each panel the large black dot and the black line indicate the planet's position and its track in the $a_\mathrm{pl}-m_\mathrm{pl}$ plane.}
         \label{fig16:reduced_migr_speed}
\end{figure}

We run several simulations with artificially increased migration times to mimic the effect of a slower inward migration. We have found that generally, a slower inward migration results in a larger final planetary mass. The reason for this result is two-fold: first, as staying longer at the planet trap in the vicinity of the pebbles accumulated, the solid core can grow larger, moreover, the growing planet spends a longer time in the disc, therefore its gaseous envelope can grow larger, too. In Figures \ref{fig16:reduced_migr_speed} one can see that the increase of the original migration timescale by a factor of two and three results in final masses as 33$M_\oplus$ (panel (b)) and 60$M_\oplus$ (panel (c)), respectively, while the final planetary mass in the non-modified timescale migration is 20$M_\oplus$, (panel (a)). We have also checked, whether the massive, 60$M_\oplus$ planet is subject to the rapid III-type migration. Since both the mass of the planet ($M_\mathrm{p}$) and the Toomre $Q$ parameter \citep{Toomre1964ApJ} depend on the distance from the star, we could plot the path of the planet in the $M_\mathrm{p} - Q$ parameter plane. In this plane, it is already identified the region, in which the runaway migration occurs, see Figure 14 in \citep{MassetPapaloizou2003ApJ}. We have found that although the planetary mass can get close, it always stays outside of the region, where the runaway migration is possible.

From Figures \ref{fig16:reduced_migr_speed}, one can also conclude that the slower inward migration also affects the overall mass of planetesimals: an increase can be seen at the onset of migration. This behaviour can be explained by the fact that the overdense ring of pebbles that follows the planet also spends more time in the given region of the disc allowing the formation of more planetesimals. At later times, this effect cannot be seen due to the faster inward migration of the more massive planet.

Our results obtained in this section suggest that a slower inward migration indeed has effects on the final mass of the planet and planetesimals formed. In future work, we will do a more detailed investigation in this direction by using more elaborated migration prescriptions \citep[e.g.][]{Paardekooper2011, Jimenez2017} in which the heating \citep[e.g.][]{Guilera2019, Guilera2021} and pebble torques are also incorporated \citep[e.g.][]{Guilera2023}.

\section{Discussion and summary}

Solar System examples, such as the main belt asteroids and Kuiper-belt objects, suggest that a large number of planetesimals should also form in the outer region of protoplanetary discs. In our work, we propose a mechanism that can trigger planetesimal formation far from the inner disc and independently of the effect of the water snowline \citep{DrazkowskaAlibert2017A&A, DrazkowskaDullemond2018A&A}. The planetesimal formation has also been studied in a static pressure bump by \citet{Guilera+2020} and \citet{Lau+2022A&A}. In the latter work, planetesimals could be dispersed in a broader region due to the gravitational interactions with the growing embryos formed in the pressure maximum. Formation of planetesimals in a broad, belt-like region has also been investigated by \citet{ShibaikeAlibert2020A&A} and \citet{ShibaikeAlibert2023planetesimal} considering an inward migrating planet. This planet opens a gas gap and similarly to our findings, in the pressure maximum at the outer edge of the gap, planetesimals form during the migration of the planet. The origin of the migrating planet, however, has not been considered in their work.

Our study show similarities to the works of \citet{Lau+2022A&A}, \citet{ShibaikeAlibert2020A&A}, and \citet{ShibaikeAlibert2023planetesimal}. These are the formation of planetesimals and a planet in a pressure trap and the formation of planetesimals in an inward-moving ring of pebbles. A significant difference is that we introduce the idea of a transient dust (and planet) trap that may develop at the boundary between two regions characterized by different accretion strengths. Due to the still limited knowledge of the exact location of such regions, we use an arbitrarily chosen boundary at 45 au, and the accretion rate of the two regions is characterised by a jump in the $\alpha$ viscosity parameter. It is very important to note that this bump in surface density diminishes as the disc evolves to a steady state. 

If the jump in the disc's viscosity is modest, no pressure maximum appears at the surface density maximum, however, in our particular cases the pressure gradient becomes very close to zero, therefore the inward drift of solids significantly slows down. Due to this slowdown, dust particles accumulate forming an overdense ring and when the conditions for the streaming instability are fulfilled, planetesimals begin to form. Since there is nothing to stop the inward drift of the ring of pebbles, it drifts more and more rapidly towards the star, because the absolute value of the (negative) pressure gradient increases as it reaches the inner region of the disc. In our particular simulations, the conditions of the streaming instability are met in a broad range of the disc, but shrinking as $\delta\alpha$ increases. Additionally, we enable the growth of a large planetesimal with mass ranging between $10^{-4}M_\oplus \leq  m_\mathrm{pl} \leq 10^{-2} M_\oplus$ when the overall mass of planetesimals exceeds 5 Moon-mass at the location of the viscosity transition. This body, depending on its initial mass, can grow up to 3 $M_\oplus$ by accreting pebbles arriving from the outer realms of the protoplanetary disc and migrating inward. 

In the case of a larger jump in the disc's viscosity, a pressure maximum also develops in connection with the surface density maximum. In the pressure maximum, the solid particles get trapped, grow, and sediment towards the disc's midplane triggering the streaming instability. Similarly to the previous cases with the modest jumps in viscosity, we also enable the growth of a large planetesimal/small protoplanet ($10^{-4}M_\oplus \leq  m_\mathrm{pl} \leq 10^{-2} M_\oplus$), when the overall mass of planetesimals exceeds 5 Moon-mass at the location of the viscosity transition. This body grows very quickly up to a planet with 10 Earth mass while partly consuming pebbles accumulated at the pressure maximum. The planet gradually opens a partial gap in the gas that traps solid particles arriving from the outer disc. These particles also form a ring of pebbles in which planetesimals form via streaming instability. As the planet trap vanishes, the planet migrates inwards, and the ring of pebbles moves together with it. In our simulations, planetesimals still form in this ring of pebbles during the planet's migration. At the end of these simulations usually a Neptune-mass planet is formed.

We have shown that a transition between an accretionally more active outer and less active inner regions promotes the formation of both planets and planetesimals. Planetesimals play an important role in planet formation being the building blocks of planets. Depending on their sizes, planetesimals can further grow either by consecutive collisions or by pebble accretion, which in turn may result in the formation of a solid core of a giant planet. In the presented scenario, the planetesimal population is not affected gravitationally by the migrating planet, this effect, which certainly results in the spreading of planetesimals in the disc will be studied in our forthcoming work. Such research may help understand the formation of the ancient population of minor bodies that evolve further collisionally and dynamically to the presently observed population of asteroids in the Solar System, and more generally, in planetary systems. 

In future work, the formation of planetary cores from the wide sea of planetesimals should also be investigated, which certainly opens a broad way for the formation of diverse planetary systems, deserving more detailed investigations.

\begin{acknowledgements}
ZsS acknowledges the support of the National Research, Development and Innovation Office (grant number MEC\_R 141396) for funding his participation in the Protostars and Planets VII conference, which significantly contributed to the research presented in this paper. OMG is partially supported by PIP-2971 from CONICET (Argentina) and by PICT 2020-03316 from Agencia I+D+i (Argentina). The authors thank the anonymous reviewer for the useful comments and suggestions that helped us considerably improve the manuscript.
\end{acknowledgements}

%
   \bibliographystyle{aa} 
   \bibliography{sandor} 

\newpage
\begin{appendix}

\section{Gas evolution equation}
To follow the gas evolution we use the diffusion-like equation presented by \cite{Lynden-BellPringle1974MNRAS}:
\begin{equation}
\frac{\partial \Sigma_\mathrm{gas}}{\partial t} = 
\frac{3}{r} \frac{\partial }{\partial r}\left[ r^{1/2} \frac{\partial }{\partial r} \left( \nu\Sigma_\mathrm{gas} r^{1/2}\right)\right],
\label{eq:gasevol}
\end{equation}
where $\Sigma_\mathrm{gas}$ is the surface density of gas, and $\nu$ is the kinematic viscosity, which according to \cite{ShakuraSunyaev1973A&A} is
approximated as $\nu = \alpha c_s^2 / \Omega_\mathrm{K}$, where $c_s = \Omega_\mathrm{K} H_\mathrm{gas}$ the sound speed, $H_\mathrm{gas}$ the disc's scale height, and $\Omega_\mathrm{K}$ the Keplerian angular velocity. The parameter $\alpha$ is dimensionless, embodying the strength of the turbulence.
To numerically solve the above partial differential equation, we provide an initial condition $\Sigma_\mathrm{gas, 0}(r)$ with appropriate boundary conditions: outflow at $r_\mathrm{min}$, and $\Sigma_\mathrm{gas}(r_\mathrm{max})=10^{-15} \mathrm{M}_\odot \mathrm{au}^{-2}$, where $r_\mathrm{max}=300$ au. Moreover, a temperature profile $T(r)$ is also provided. 

\section{N-Body integration including forces from the ambient disc}
To solve the N-body problem, in which the planetary bodies are embedded in the protoplanetary disc, besides the mutual gravitational forces, forces arising from the ambient disc should also be calculated. The force (per unit mass) acting on the $i$th planet is given by the following system of differential equations \citep{CresswellNelson2008A&A}:
\begin{equation}
\vec{\ddot r}_i = \vec{\ddot r}_{i,\mathrm{grav}} - \frac{\vec{\dot r}_i}{2\tau_{a_i}} - 
\frac{ 2(\vec{\dot r}_i \cdot \vec{r}_i) \vec{r}_i} {r^2_i \tau_{e_i}} - 
\frac{(\vec{\dot r}_i \cdot \vec{k})\vec{k}}{\tau_{inc_i}},
\label{parametrized_nbody}
\end{equation}
where 
$\vec{\ddot r}_{i,\mathrm{grav}}$ the gravitational force, $\vec{r}$ and $\vec{\dot r}$ are the position and velocity vector of the $i$th planet, while $\tau_{a_i}$ is the migration time, $\tau_{e_i}$ and $\tau_{inc_i}$ are the eccentricity and inclination damping times, respectively and $\vec{k}$ is the unit vector in z-direction. The migration and damping timescales are calculated based on the prescription of \citet{Tanaka+2002ApJ} and \citet{TanakaWard2004ApJ} for the local isothermal case using the gradients of the gas surface density and temperature. The N-body integration is done using a Bulirsch-Stoer scheme \citep{Press+1989numrec} that we developed and proved to be efficient and accurate in our earlier investigations \citep{Sandor+2011ApJ, Horn+2012ApJ}.

\section{Planet growth by pebble accretion}
In our model, pebble accretion is incorporated following the work of \citet{Venturini+2020A&Aa, Venturini+2020A&Ab}. One can distinguish between 2D and 3D accretion. Pebble accretion is in 2D if pebbles are more or less sedimented towards the disc's midplane and happens for low $\alpha$ values or high Stokes-numbers. In this case, when $\mathrm{St}<0.1$, the pebble accretion rate of a growing planetary core is \citep{LambrechtsJohansen2014A&A}:
\begin{equation}
\dot M_\mathrm{peb,2D} = 2 \left(\frac{\mathrm{St}}{0.1}\right)^{2/3}R_\mathrm{Hill}v_\mathrm{Hill}\Sigma_\mathrm{peb},
\label{eq:pebble_accr_rate2D}
\end{equation}
where $R_\mathrm{Hill}$ is the Hill radius of the planet, and $v_\mathrm{Hill}$ is the Hill velocity, e.g. the orbital velocity around the planetary core at the Hill radius. If $0.1 < \mathrm{St} < 1$, the above expression should be evaluated with $\mathrm{St=0.1}$. If due to the stronger turbulence pebbles are stirred up from the disc's midplane, pebble accretion happens in 3D. The condition for 3D accretion is:
\begin{equation}
f_\mathrm{3D} := \frac{1}{2}\sqrt{\frac{\pi}{2}}\left(\frac{\mathrm{St}}{0.1}\right)^{1/3}\frac{R_\mathrm{Hill}}{H_\mathrm{peb}} < 1,
\end{equation}
where
\begin{equation}
    H_\mathrm{peb}=H_\mathrm{gas}\sqrt{\frac{\alpha}{\alpha+\mathrm{St}}},
\end{equation}
is the scale height of pebbles, given by \citet{YoudinLithwick2007Icar}. In this case, the pebble accretion rate is 
\begin{equation}
\dot M_\mathrm{3D,peb} = f_\mathrm{3D} \dot M_\mathrm{peb,2D}.
\label{eq:pebble_accr_rate3D}
\end{equation}

\section{Gap opening by the growing planet}
A growing core exerts a torque on the ambient gaseous material in its region of corotation pushing gas away. On the other hand, due to the turbulent viscosity gas is replenished at a certain level. As a net effect, planetary cores more massive than a few times the Earth's mass open a partial gap that is almost fully cleared if the planet reaches the Jupiter-mass regime. In our model gap opening is modelled by the analytic formula provided by \citet{Chambers2021ApJ}, which is based on the work of \citet{Kanagawa+2018ApJ}. According to the latter, the depth of the gap is given by
\begin{equation}
F_\mathrm{gap} = \frac{1}{1+0.04K},
\end{equation} 
while its width is
\begin{equation}
w_\mathrm{gap} = \frac{a_\mathrm{pl}}{4}\left(\frac{m_\mathrm{pl}}{M_*}\right)^{1/2}\left(\frac{a_\mathrm{pl}}{H_\mathrm{gas}}\right)^{3/4}\left(\frac{1}{\alpha}\right)^{1/4},
\end{equation}
where $a_\mathrm{pl}$ and $m_\mathrm{pl}$ are the semi-major axes and mass of the growing planet, respectively, $H_\mathrm{gas}$ is the scale height of the gas, $M_*$ is the mass of the star, and 
\begin{equation}
K = \left(\frac{m_\mathrm{pl}}{M_*}\right)^2\left(\frac{a_\mathrm{pl}}{H_\mathrm{gas}}\right)^5\frac{1}{\alpha}.
\end{equation}
Having defined the depth and the width of the gap, the surface density profile $\Sigma_\mathrm{gas}(a)$ is given by
\begin{equation}
\frac{\Sigma_\mathrm{gas}(a)}{\Sigma_\mathrm{un,gap}(a)} = 1-(1-F_\mathrm{gap})\exp\left[-\frac{1}{4}\left(\frac{a-a_p}{w_\mathrm{gap}}\right)^4 \right],
\label{eq:gapopening}
\end{equation}
where $\Sigma_\mathrm{un,gap}(a)$ is the unperturbed surface density profile around the position of the planetary core. This profile approximately mimics the shape of the gap calculated by \citet{Duffell2020ApJ}.

\section{Dust evolution and dynamics}
\label{subsec:dustevolmodel}
In protoplanetary discs, the solid-to-gas mass ratio (or metallicity) is assumed initially to be small ($\varepsilon = 0.01$), therefore dust transport can be described by an advection-diffusion equation. In our simulations, we use the two-population model of \citet{Birnstiel+2012A&A} for the evolution of the dust surface density, in which the solid material is treated as large grains and $\mu$m-sized dust particles. The overall transport of dust particles can be investigated by solving the transport equation 
\begin{equation}
\begin{split}
 \frac{\partial \Sigma_\mathrm{dust}}{\partial t}  & +  \frac{1}{r}\frac{\partial}{\partial r}\left( r\Sigma_\mathrm{dust} \tilde u \right) - \\
& \frac{1}{r}\frac{\partial}{\partial r}\left[r \Sigma_\mathrm{gas} D^*\frac{\partial}{\partial r}\left(\frac{\Sigma_\mathrm{dust}}{\Sigma_\mathrm{gas}}\right)\right] - \mathcal{S}_\mathrm{acc}(r,t) - \mathcal{S}_\mathrm{plts}(r,t)= 0,
\end{split}
\label{eq:advection}
\end{equation}
for $\Sigma_\mathrm{dust}$ being the surface density of dust particles, with the mass-weighted radial drift velocity for advection $\tilde u$ and diffusion coefficient $D^*$ whose definitions can be found in \citet{Birnstiel+2012A&A}. The third term $\mathcal{S}(r,t)_\mathrm{acc}$ is a sink term due to the pebble accretion by the growing planetary core, while the fourth term, $\mathcal{S}(r,t)_\mathrm{plts}$ takes into account the loss of pebbles due to the streaming instability. We note that for $\Sigma_\mathrm{dust}$ similar initial and boundary conditions are applied as for $\Sigma_\mathrm{gas}$, apart from the multiplicative factor $\epsilon = 0.01$.

The radial velocity $u$ of a dust particle depends mainly on its size and the radial pressure gradient of gas \citep{Weidenschilling1977MNRAS}:
\begin{equation}
u \approx \frac{2}{\left(\mathrm{St} + 
\mathrm{St}^{-1}\right)} \frac{c_s^2}{2 V_\mathrm{K}}\frac{\mathrm{d}\ln P}{\mathrm{d}\ln r},
\label{eq:drift_velocity}
\end{equation}
where $V_\mathrm{K}$ is the circular Keplerian velocity, $P$ is the pressure of the gas, and $c_s$ is the sound speed of the gas at the given location. The Stokes-number (dimensionless friction time) $\mathrm{St}$ of a dust particle with radius $s$ and physical density $\rho_p$ can be easily calculated both in the Epstein or Stokes/quadratic drag regimes \citep{Rafikov2004AJ, Chambers2008Icar}. The gas pressure $P$ in the disc's midplane 
can be calculated as: 
\begin{equation}
P = \rho_\mathrm{gas} c_s^2 = \frac{\Sigma_\mathrm{gas} \Omega_\mathrm{K} c_s}{\sqrt{2\pi}}.
\label{eq:pressure}
\end{equation}

Initially dust particles with size $s_0$ (set to 1 $\mu$m in our simulations) grow by coagulation on the timescale
\begin{equation}
\tau_\mathrm{growth} \approx \frac{1}{\varepsilon \Omega_\mathrm{K}},
\label{eq:growtau}
\end{equation}
until reaching the maximum value
\begin{equation}
s_\mathrm{max}(r,t) = \min \left[s_\mathrm{frag}, s_\mathrm{drift}, s_\mathrm{df}, s_0 \exp\left(\frac{t}{\tau_\mathrm{growth}}\right)\right],
\label{eq:particle_growth}
\end{equation}
where
$s_\mathrm{frag}$ is the fragmentation size limit, depending on the fragmentation threshold velocity of dust aggregates $u_\mathrm{frag}$ and the turbulence of gas, $s_\mathrm{drift}$ is the dust size limit due to the fast radial drift of particles, and $s_\mathrm{df}$ is the dust size limit due to the radial drift-induced fragmentation.
see \citet{Birnstiel+2012A&A} for more details. It is commonly assumed that the fragmentation threshold velocity $u_\mathrm{frag}$ of dust particles varies between $1-10$ m/s for aggregates depending on whether they are formed by silicates (lowest value) or ices (highest value), for example \citet{GundlachBlum2015ApJ}. 

\section{Planetesimal formation via streaming instability}
Since the pressure maximum is formed inside an accretionally less active region, characterized by relatively low values of $\alpha = 10^{-4} - 10^{-3}$, dust particles can grow quite fast until they are almost sedimented in the disc's midplane and the (volumetric) density of pebbles, $\rho_\mathrm{dust}$, is comparable to that of the gas $\rho_\mathrm{gas}$. Analogously to the density of gas in the disc's midplane, 
\begin{equation}
  \rho_\mathrm{dust} = \frac{\Sigma_\mathrm{dust}}{\sqrt{2\pi} H_\mathrm{dust}},
\end{equation}
where the scale height of dust $H_\mathrm{dust}$, in the presence of a turbulence characterized by the parameter $\alpha$, is given by \citet{YoudinLithwick2007Icar} as
\begin{equation}
  H_\mathrm{dust} = H_\mathrm{gas}\sqrt{\frac{\alpha}{\alpha+\mathrm{St}}},
\end{equation}
being $H_\mathrm{gas}$ the scale height of gas.
The condition of the onset of streaming instability is fulfilled when
\begin{equation}
\frac{\rho_\mathrm{dust}}{\rho_\mathrm{gas}} = \frac{\Sigma_\mathrm{dust}}{\Sigma_\mathrm{gas}}\sqrt{\frac{\alpha+\mathrm{St}}{\alpha}} > 1, 
\label{eq:streaming_cond}
\end{equation}
for particles with Stokes number $\mathrm{St} > 0.01$ \citep{Drazkowska+2016A&A}. We note that a recent study \cite{LiYoudin2021ApJ} found a less restrictive condition for the streaming instability, however in our work, we still use the above, more conservative approach.
Whenever the condition given by Equation \eqref{eq:streaming_cond} holds, pebbles accumulated in the pressure maximum are transformed to large planetesimals (assumed to be insensitive to drag force) with the rate: 
\begin{equation}
\dot \Sigma_\mathrm{pl}(r) = \zeta\frac{\Sigma_\mathrm{dust}(r)}{T_\mathrm{K}},
\label{eq:planetesimalformationeficiency}
\end{equation} 
where $\zeta$ is the efficiency parameter of planetesimal formation and $T_\mathrm{K} = 2\pi\Omega_\mathrm{K}^{-1}$ is the Keplerian orbital period at $r$. 

\section{Onset of giant planet formation and accretion of a gaseous envelope}
A planetary core can maintain a massive gaseous envelope in hydrostatic equilibrium if there is a mechanism that assures a continuous heat release at the core's surface. Assuming that the gaseous envelope is optically thick, the increased temperature supplies the extra pressure that counteracts the envelope's gravity. The heat release is due to pebble or planetesimal accretion when the gravitational potential energy of solid material is converted to heat when reaching the core's surface. If the envelope becomes massive enough at a given solid accretion rate, this extra pressure cannot counterbalance gravity and the envelope will collapse on a Kelvin-Helmholtz timescale leading to the formation of a gas giant planet. According to the numerical integrations of the equations for evolution and structure, \citet{Ikoma+2000ApJ}  find the critical core mass until the gaseous envelope is in hydrostatic equilibrium. If the mass of the core is bigger than this critical mass, the gaseous envelope collapses and a giant planet forms. The critical core mass is given by:
\begin{equation}
M_\mathrm{core,cr} \approx 7\left(\frac{\dot M_\mathrm{core}}{1\times 10^{-7} M_\oplus \mathrm{yr}^{-1}}\right)^q\left(\frac{\kappa_\mathrm{gr}}{1\mathrm{cm}^2 \mathrm{g}^{-1}}\right)^s,
\label{eq:critical_coremass}
\end{equation}  
depending on the accretion rate of solids $\dot M_\mathrm{core}$ and the dust opacity $\kappa_\mathrm{gr}$, while the exponents $q$ and $s$ range between 0.2-0.3. We note that in the outer disc $\kappa_\mathrm{gr} = 1 \mathrm{g/cm}^2$ \citep[see][]{Guilera+2020,Venturini+2020A&Aa}, thus in our simulations the exponent $s$ does not enter in the formula of the critical core mass.

Once the mass of the planetary core becomes larger than the critical mass $m_\mathrm{pl} > M_\mathrm{core,cr}$, the gaseous envelope begins to collapse on a Kelvin-Helmholtz timescale $\tau_\mathrm{env}$ that depends on the core's mass and the grain opacity $\kappa_\mathrm{gr}$ as given by \citet{Ikoma+2000ApJ}:
\begin{equation}
  \tau_\mathrm{env} = 10^8 \left(\frac{M_\mathrm{core,cr}}{M_\oplus} \right)^{-2.5} \left( \frac{\kappa_\mathrm{gr}}{\mathrm{cm^2/yr}}\right) \text{yr}.
  \label{eq:KH_time}
\end{equation}
The envelope's growth is then governed by $\tau_\mathrm{env}$, as 
\begin{equation}
    \dot M_\mathrm{env} = \frac{M_\mathrm{env}}{\tau_\mathrm{env}}.
\end{equation}
The initial condition to the above differential equation is $M_\mathrm{env}^0=M_\mathrm{env}(t_0)$, where $t_0$ is the time when the collapse of the envelope begins. With this initial condition, the envelope's growth is purely exponential:
\begin{equation}
M_\mathrm{env}(t) =  M_\mathrm{env}^0 \exp\left(\frac{t-t_0}{\tau_\mathrm{env}}\right).
\label{eq:envelope_growth}
\end{equation}

\section{Dependence on the initial mass of the inserted massive planetesimal}
We present our results on how the mass of the large planetesimal affects the mass and migration history of the planet formed. We use three mass values for the planetesimal inserted in the pressure trap, namely $10^{-4} M_\oplus$, $3\cdot 10^{-3} M_\oplus$, and $10^{-2} M_\oplus$, see our detailed explanation in Section \ref{subsec:grembryo}. We assume that such bodies are indeed formed in the pressure traps developed in our simulations. We have also checked that even the body with the smallest mass ($10^{-4} M_\oplus$) is in the Hill regime of pebble accretion as its mass is above the transition mass between Bondi and Hill regimes. 

\begin{figure}[]
   \centering
    \includegraphics[width=\columnwidth]{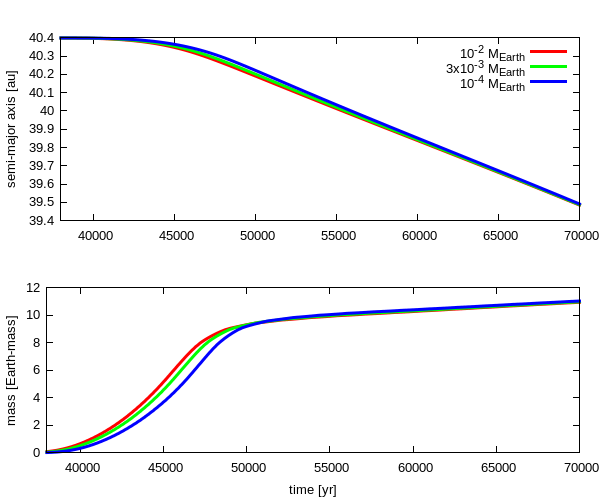}
      \caption{Migration (upper panel) and the mass growth (bottom panel) of the planet in the three simulations using three different initial masses, see the text for the details. The curves are coloured corresponding to the initial masses.}
      \label{fig:embryo_diff}
\end{figure}

First, we show our results when the dust traps form in transient pressure maxima ($\delta\alpha \leq 0.15$). In these cases, the large planetesimal stays long enough in a pebble-rich region, thus the initial mass difference disappears in a very short time. Results for $\delta\alpha = 0.15$ are shown in Figure \ref{fig:embryo_diff} in which the mass growth (bottom panel) and migration history (upper panel) of the planet formed are almost indistinguishable for simulations with different initial masses.
\begin{figure}[]
   \centering
    \includegraphics[width=\columnwidth]{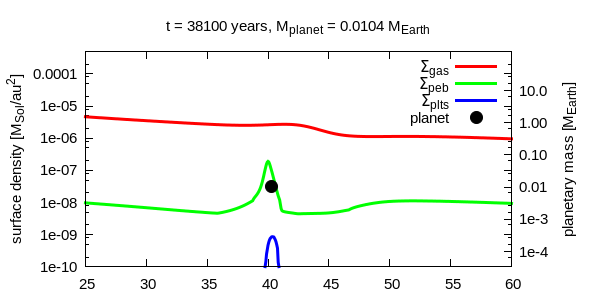}
    \includegraphics[width=\columnwidth]{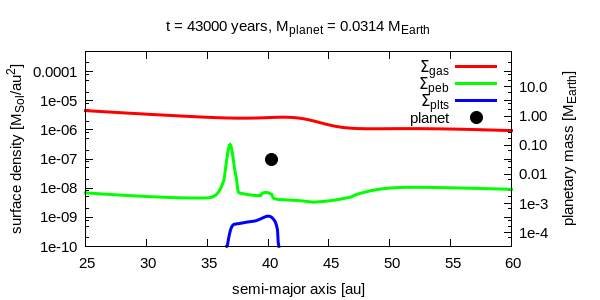}
      \caption{The ring of pebbles rapidly drifts away from the formation place of the large planetesimal, denoted by the black dot.}
      \label{fig:embryo_diff2}
\end{figure}

In the case when $\delta\alpha \geq 0.2$ no pressure maximum develops, therefore the ring of pebbles is continuously drifting inward. As a consequence, planetesimals formed are exposed to the high pebble flux for a significantly shorter time than in the cases $\delta\alpha \leq 0.15$, see Figure  \ref{fig:embryo_diff2}, therefore their initial mass differences do not disappear. As a consequence, smaller mass planetesimals grow to smaller mass planets, too.

\end{appendix}
\end{document}